\documentclass[aps,reprint,superscriptaddress,longbibliography]{revtex4-1}
\pdfpageattr {/Group << /S /Transparency /I true /CS /DeviceRGB>>}
\usepackage{amsmath}
\usepackage{amsthm}
\usepackage{physics}
\usepackage{amsfonts}
\usepackage{graphicx}
\usepackage{wrapfig}
\usepackage{times,txfonts}
\usepackage[colorlinks=true,linkcolor=blue,urlcolor=blue,citecolor=blue,pdfusetitle]{hyperref}
\usepackage{hyperref}
\usepackage[dvipsnames]{xcolor}

\renewcommand{\ket}[1]{\ensuremath{|#1\rangle}}

\begin{document}
	\title{Multiparameter critical quantum metrology with impurity probes}
	
	\author{George Mihailescu}
	\email[]{george.mihailescu@ucdconnect.ie}
	\affiliation{School of Physics, University College Dublin, Belfield, Dublin 4, Ireland}
	\affiliation{Centre for Quantum Engineering, Science, and Technology, University College Dublin,  Dublin 4, Ireland}
        \author{Abolfazl Bayat}
        \email[]{abolfazl.bayat@uestc.edu.cn}
        \affiliation{Institute of Fundamental and Frontier Sciences, University of Electronic Science and Technology of China, Chengdu 611731, China}
        \affiliation{Key Laboratory of Quantum Physics and Photonic Quantum Information, Ministry of Education,
University of Electronic Science and Technology of China, Chengdu 611731, China}

	\author{Steve Campbell}
	\email[]{steve.campbell@ucd.ie }
	\affiliation{School of Physics, University College Dublin, Belfield, Dublin 4, Ireland}
	\affiliation{Centre for Quantum Engineering, Science, and Technology, University College Dublin, Dublin 4, Ireland}
        \affiliation{Dahlem Center for Complex Quantum Systems, Freie Universit\"at Berlin, Arnimallee 14, 14195 Berlin, Germany}
	\author{Andrew K. Mitchell}
	\email[]{andrew.mitchell@ucd.ie}
	\affiliation{School of Physics, University College Dublin, Belfield, Dublin 4, Ireland}
	\affiliation{Centre for Quantum Engineering, Science, and Technology, University College Dublin,  Dublin 4, Ireland}
	

\begin{abstract}\noindent
Quantum systems can be used as probes in the context of metrology for enhanced parameter estimation. In particular, the delicacy of \textit{critical} systems to perturbations can make them ideal sensors. 
Arguably the simplest realistic probe system is a spin-$\tfrac{1}{2}$ impurity, which can be manipulated and measured \textit{in-situ} when embedded in a fermionic environment. 
Although entanglement between a single impurity probe and its environment produces nontrivial many-body effects, criticality cannot be leveraged for sensing.
Here we introduce instead the two-impurity Kondo (2IK) model as a novel paradigm for critical quantum metrology, and examine the multiparameter estimation scenario at finite temperature. We explore the full metrological phase diagram numerically and obtain exact analytic results near criticality. 
Enhanced sensitivity to the inter-impurity coupling driving a second-order phase transition is evidenced by diverging quantum Fisher information (QFI) and quantum signal-to-noise ratio (QSNR). 
However, with uncertainty in both coupling strength and temperature, the multiparameter QFI matrix becomes singular -- even though the parameters to be estimated are independent -- resulting in vanishing QSNRs. We demonstrate that by applying a known control field, the singularity can be removed and measurement sensitivity restored. For general systems, we show that the degradation in the QSNR due to uncertainties in another parameter is controlled by the degree of correlation between the unknown parameters.    
\end{abstract}
\maketitle
	
\section{Introduction}

Quantum probes are known for their advantage over classical sensors in achieving higher precision using the same resources~\cite{degen2017quantum}.  Originally, such advantage has been accomplished through exploiting special forms of quantum superposition, known as GHZ states~\cite{greenberger1989going}, to reach quadratic improvement in the precision of detecting external signals~\cite{giovannetti2004quantum,leibfried2004toward}. However, GHZ-based quantum sensors are hard to scale up and are prone to decoherence~\cite{demkowicz2012elusive,zhou2024limits} and perturbation~\cite{de2013quantum}. Alternatively, strongly correlated quantum many-body systems near their phase transitions have been identified as a resource for quantum enhanced sensitivity~\cite{venuti2007quantum,schwandt2009quantum,albuquerque2010quantum,gritsev2009universal,
gu2008fidelity,greschner2013fidelity,frerot2018quantum,zhou2020quantum,rams2018limits,chu2021dynamic, di2022multiparameter,chu2023strong,salvia2023critical,rodríguez2023strongly,bressanini2024multiparameter,cavazzoni2024characterization}. Several kinds of phase transitions have been proposed for quantum sensing, including second-order~\cite{zanardi2006ground,abasto2008fidelity,sun2010fisher,zanardi2008quantum,damski2013exact,salvatori2014quantum,yang2022super,fernandez2018heisenberg,montenegro2022sequential,ozaydin2015quantum, garbe2022critical,mirkhalaf2021criticality,Saulo,Karol2,Karol1,lü2024critical,Simone}, topological~\cite{budich2020nonhermitian,koch2022quantum,free2022sarkar,zhang2023topological,Yang_2024}, superradiant and Rabi type ~\cite{bin2019mass,dicandia2023critical,garbe2020critical, heugel2019quantum, lorenzo2017quantum,wu2021criticality,ying2022critical,tang2023enhancement,zhu2023rabi,garbe2022exponential},  dynamical~\cite{tsang2013quantum,macieszczak2016dynamical,carollo2018uhlmann}, Floquet~\cite{lang2015dynamical,mishra2021driving,mishra2022integrable}, continuous environmental monitoring~\cite{ilias2022criticality,boeyens2023probe}, Stark localization ~\cite{he22023stark, Yousefjani2023long}, disorder-induced ~\cite{bhattacharyya2023disorderinduced,sahoo2023localization}, and boundary time crystals~\cite{montenegro2023boundary,cabot2023continuous}. Furthermore, certain criticality-based sensing mechanisms have been experimentally realized in NV-centers in diamond~\cite{yu2022experimental}, NMR~\cite{liu2021experimental}, trapped ions~\cite{ilias2023criticalityenhanced}, and Rydberg atoms~\cite{ding2022enhanced}.
Although quantum phase transitions strictly occur at zero temperature, in practice any physical realization will be performed at finite temperatures where thermal fluctuations become important. To further complicate matters, the temperature of the system itself might not be known precisely. As a consequence, there has been a growing interest in developing thermometric schemes that exploit quantum systems~\cite{correa2015individual, Paris_Landau_Percision_Thermo,Quantum_Limits_Thermo,MitchisonPRL,Campbell_Percision_Thermo_QSL, Brattegard2023, BrenesThermometry,yu2023criticalityenhanced,yang2024sequential,srivastava2023topological,verma2024measuring}. These issues give rise to several questions, including: (i) How does nonzero temperature affect the sensitivity of a criticality-based quantum sensor? (ii) Can quantum criticality also boost the sensitivity of temperature estimation?

Perhaps the most well-known theorem in metrology is the Cram\'er-Rao inequality \cite{Paris_Quantum_Estimation} which puts a fundamental bound on the uncertainty of inferring one or more unknown parameters. Let us consider a quantum probe which encodes  $n$ parameters $\vec{\lambda} =(\lambda_1,~\lambda_2,~...,~\lambda_n)^T$ in its density matrix $\varrho(\vec{\lambda})$. The uncertainty in estimating the parameters $\vec{\lambda}$, through performing a suitable measurement on the probe, can be quantified by the covariance matrix $\textbf{Cov}[\vec{\lambda},\vec{\lambda}]$ with components given by $\text{Cov}(\lambda_i,\lambda_j)=\langle (\lambda_i-\langle \lambda_i\rangle )(\lambda_j-\langle \lambda_j\rangle )\rangle$. The individual variances are then simply the diagonal elements, $\text{Var}(\lambda_i)\equiv \text{Cov}(\lambda_i,\lambda_i)$. The Cram\'er-Rao inequality is given by,
\begin{equation}
\label{eq:MPCRB}
\textbf{Cov}\left[\vec{\lambda}\right] \geq \frac{1}{N} \textbf{F}^{-1}
\end{equation}
where $N$ is the number of samples and $\textbf{F}$ is the Fisher information matrix~\cite{helstrom1969quantum,Holevo}. A crucial point is that the Cram\'er-Rao bound is only meaningful when the Fisher information matrix is invertible. For example, in a two parameter setting, if the parameters are not independent from each other -- that is, they can can be rescaled to a single effective parameter, one can show that the Fisher information matrix becomes singular and thus non-invertible. However, this is only a necessary condition and there might be other situations which result in a non-invertible Fisher information matrix.  A potential scheme for such a problem is in  criticality-based quantum sensing where the probe operates at a nonzero temperature that may not be known with certainty. In this case, one has to address the problem through a multiparameter sensing analysis, where temperature is also treated as an unknown parameter. In principle, the unknown microscopic parameter which drives the phase transition is independent from the temperature of the system. However, the way that these parameters are encoded in the quantum state of the probe may nonetheless result in a singular Fisher information matrix -- especially in systems constrained by high symmetries or near quantum critical points where the low-energy physics exhibits emergent single-parameter scaling \cite{sachdev1999quantum}. In such cases, can one make the Fisher information matrix invertible again, thereby allowing for an effective sensing protocol?

In this work we consider the two-impurity Kondo (2IK) model \cite{jayaprakash1981two,jones1988low,affleck1992exact,affleck1995conformal,mitchell2012two,sela2011exact,andrew_entropy_exact}, a famous paradigm in condensed matter physics for quantum criticality in a strongly-correlated many-body system. Originally conceived to describe the through-lattice (RKKY) coupling between two magnetic impurities (such as iron atoms) embedded in a host metal (such as gold), the 2IK quantum phase transition also captures the essence of the competition between magnetic ordering and heavy-fermion physics in real correlated materials \cite{hewson1997kondo}. The model can be realized in semiconductor quantum dot devices \cite{van2002electron,izumida2000two,zarand2006quantum,jeong2001kondo}, and indeed a closely related variant of the 2IK critical point was observed experimentally very recently in Ref.~\cite{pouse2023quantum,*karki2023z}. The 2IK model features two exchange-coupled spin-$\tfrac{1}{2}$ ``impurity" qubits, each coupled to its own fermionic environment (taken to be metallic continua of conduction electrons). 
Although such impurities are non-invasive in the sense that impurity effects in bulk systems are always non-extensive, they can still induce \textit{local} criticality (also known as boundary critical phenomena) \cite{affleck1992exact}. In the 2IK model, a nontrivial critical point, obtained by tuning the inter-impurity coupling strength, separates a phase in which the impurities bind together into a local spin-singlet state, from a phase in which each impurity is separately screened by conduction electrons through the Kondo effect \cite{affleck1992exact,mitchell2012two}. The underlying physics is controlled by the development of strong many-body entanglement between the probe impurities and the electronic environment \cite{bayat2012entanglement,bayat2014order}.  
The presence of a quantum phase transition whose critical properties survive thermal fluctuations (up to the so-called Kondo temperature) makes the 2IK model an excellent testbed for studying the performance of a critical sensor at nonzero temperature. 
Furthermore, we note that such interacting quantum condensed matter systems in the thermodynamic limit can in practice be straightforwardly tuned
into their critical regimes at finite temperatures, without the need for meticulous and costly critical ground state preparation techniques \cite{Gietka2021adiabaticcritical,PhysRevX.8.021022}, as demonstrated experimentally in nanoelectronics device realizations \cite{potok2007observation,iftikhar2018tunable,pouse2023quantum}.

We regard the two coupled impurities in the 2IK model as a metrological probe, whose reduced state is characterized by the impurity singlet fraction (spin-spin correlator) -- an experimentally relevant physical observable. The additional internal structure of the coupled impurity probe relative to a single impurity probe \cite{Mihailescu_Thermometry} endows a far richer metrological phase diagram. We consider estimation of either the environment temperature $T$ or the inter-probe coupling $K$, as well as the arguably more realistic multiparameter estimation scenario where neither $T$ nor $K$ are known with absolute certainty. In the latter scenario, we find that the quantum Fisher information matrix (QFIM) becomes singular -- despite temperature and impurity coupling being independent parameters. Such a situation prohibits inference of either parameter, and the corresponding quantum signal to noise ratio (QSNR) for both parameters vanishes. We propose a strategy to remedy this: by applying a known control field, the QFIM singularity is removed and the ability to perform multiparameter estimation is restored.

The paper is organised as follows: In Sec.~\ref{sec2:Param_Estimation} we briefly recapitulate the fundamentals of multiparameter estimation theory, introducing the QFIM and Cram\'er-Rao bound. We identify the QSNR as the key figure of merit, and introduce a novel generalization of this quantity in the multiparametric setting. In particular, we show that 
the QSNR for a given parameter is always reduced by uncertainties in another parameter, with the degree of 
degradation controlled by the degree of \textit{correlation} between the unknown parameters. In Sec.~\ref{sec3:model} we introduce the 2IK model and contextualize the physical regimes in which we expect to observe quantum critical features. We take the two spin-$\tfrac{1}{2}$ exchange coupled impurity qubits as our metrological probe, and show how parameter estimation sensitivity can be extracted from the spin-spin correlation function, an experimentally-motivated physical observable. In Sec.~\ref{Sec:large_K} we provide analytical results for single parameter estimation in the simple but instructive limit of large inter-probe coupling strength. In Sec.~\ref{Sec:NRG_Results} we investigate single-parameter metrology in the full many-body system, obtaining numerically exact results using the Numerical Renormalization Group (NRG) technique \cite{wilson1975renormalization,*bulla2008numerical}, showing how strong correlations and quantum criticality affect sensing capabilities in the 2IK system.  In Sec.~\ref{sec:exact_results} we derive closed-form exact analytic results for the QSNRs in the vicinity of the quantum critical point. Our solution is obtained by relating the QFI to changes in the probe entropy, and constitutes a rare example in which exact results can be obtained for an interacting quantum many-body system at finite temperatures, near a nontrivial second-order quantum phase transition. In Sec.~\ref{Sec:MultiParam_Background} we consider explicitly the multiparameter scenario. Here we explore the QFIM singularity that arises when we have uncertainty in both system temperature $T$ and probe coupling strength $K$. The singularity in the QFIM that prevents multiparameter estimation is shown to be connected to the SU(2) spin symmetry of the probe reduced state. We further demonstrate that by applying a known control field that breaks this symmetry, the singularity is removed and multiparameter estimation sensitivity is restored. Our results indicate a dramatic difference in the effectiveness of critical quantum sensing when uncertainty in more than one parameter is taken into account. This is an essential practical consideration since any experiment must be performed at finite temperature, and there is typically some experimental uncertainty in determining this temperature.


\section{Parameter Estimation}\label{sec2:Param_Estimation}
We begin by introducing the tools necessary for multiparameter estimation, in particular, the quantum Fisher information and the Cram\'er-Rao bound (CRB)~\cite{Paris_Quantum_Estimation,toth2014quantum,Liu_2020_QFIM_MPE_Review}. In what follows, we present the formalism for the multiparameter setting for generality. However we emphasize that the single parameter estimation scenario corresponds to the special case where all parameters except the one to be estimated are assumed to be known with certainty. We will extensively discuss single parameter estimation in Secs.~\ref{sec3:model}-\ref{sec:exact_results} and consider explicitly the multiparameter case in Sec.~\ref{Sec:MultiParam_Background}.

The Fisher information matrix appearing in Eq.~\eqref{eq:MPCRB} is an $n\times n$ positive semi-definite matrix for a system with $n$ unknown parameters $\vec{\lambda} =(\lambda_1,~\lambda_2,~...,~\lambda_n)^T$. Its elements are given by,
\begin{equation}
    \label{Eq:Multi_Param_FI_Def}
	\textbf{F}_{i,j} = \mathbb{E}\left[ \left( \partial_{\lambda_i}\ln{p\left(x_k \mid \lambda_i\right)}\right)\left( \partial_{\lambda_j}\ln{p\left(x_k \mid \lambda_j\right)}\right)\right]
\end{equation}
where $p(x_k | \lambda_i)$ denotes the conditional probability of obtaining outcome $x_k$ given the parameter has value $\lambda_i$~\footnote{We remark that, in the limiting case, when we have a single parameter, $\lambda$, that we wish to infer, the Fisher information in Eq.~\eqref{Eq:Multi_Param_FI_Def} reduces to $F\left(\lambda\right) = E_\lambda\left[ \partial_\lambda \ln{p\left(x_k \mid \lambda\right)} \right]^2$ and the CRB becomes $\text{Var}\left(\lambda\right) \geq {\left(NF\right)}^{-1}$, establishing a lower-bound on the mean square error, $\text{Var}\left(\lambda\right) = E_\lambda \left[\left( \hat{\lambda}\left( \{ x \}\right) - \lambda\right)^2 \right]$, of any estimator of the parameter $\lambda$.}.

In the quantum setting \cite{Paris_Quantum_Estimation}, we consider parameter-encoded quantum states, $\hat{\varrho}(\vec{\lambda})$, whose measured outcome value, $x_k$, can be obtained through a set of positive operator-valued measurements (POVMs) denoted $\{ \Pi_i \}$. The parameter dependent conditional probability of these outcomes, defined through the Born rule $p\left(x_k \mid \lambda_i\right) = \Tr{\Pi_i~\hat{\varrho}(\vec{\lambda})}$, allows for the construction of unbiased estimators. The \textit{quantum} Fisher information matrix (QFIM) $\boldsymbol{\mathcal{H}}$, is obtained through an optimization over all possible measurements. Its elements are defined in terms of the symmetric logarithmic derivative (SLD) operators as,
\begin{equation}
    \label{eq:QFIM}
	\boldsymbol{\mathcal{H}}_{\lambda_i,\lambda_j} = \frac{1}{2}\text{Tr}\left(\hat{\varrho}\acomm{\hat{L}_i}{\hat{L}_j}\right)
\end{equation}
where $\hat{L}_i$ correspond to the ideal measurement for parameter $\lambda_i$. Formally, the SLD is defined by the solution of the self-adjoint operator equation $\partial_{\lambda_i} \hat{\varrho}(\vec{\lambda}) = \frac{1}{2}\left(L_i \hat{\varrho}(\vec{\lambda}) + \hat{\varrho}(\vec{\lambda}) L_i \right)$. As such, we note that the diagonal elements of the QFIM are identical to the single-parameter QFI, that is $\boldsymbol{\mathcal{H}}_{\lambda_i,\lambda_i}=\mathcal{H}(\lambda_i)$.

The quantum multiparmater CRB reads \cite{Holevo},
\begin{equation}
\label{eq:MPQCRB}
\textbf{Cov}\left[\vec{\lambda}\right] \geq \frac{1}{N}\boldsymbol{\mathcal{H}}^{-1} \;,
\end{equation}
where we set $N\!=\!1$ for the single-shot measurement case considered hereafter, 
and we remark that the bound holds element-wise in this matrix inequality. While the diagonal elements of the QFIM, $\boldsymbol{\mathcal{H}}_{\lambda_i,\lambda_i}$, on their own provide information only about the measurement precision for parameter $\lambda_i$ in the single-parameter estimation scenario, the matrix inverse operation in Eq.~\eqref{eq:MPQCRB} means that the other elements of the QFIM affect measurement precision when we have uncertainty in any of the other parameters. 
Indeed, as shown in Appendix~\ref{Appendix:Uncertainties_multiparameter} and discussed further below, the precision of estimating a given parameter is always \textit{reduced} by uncertainties in other parameters.  In addition, the bounds for all parameters may not be simultaneously saturable using a single measurement since the SLD operators $\hat{L}_i$ for different parameters $\lambda_i$ may be incompatible~\cite{Nat_Incomp,heinosaari2016invitation,candeloro2024dimension}. Thus, a single optimal measurement basis shared by all the parameters might not exist, in which case precision trade-offs in the multiparameter estimation problem are unavoidable at the fundamental level.

It is important to establish how well one can distinguish the inferred parameter signal from the measurement noise. For example, in situations where the QFI indicates a region of high precision but the signal itself is extremely small in this region, accurate parameter estimation remains challenging in practice. For this reason, we focus on the QSNR
\begin{equation}\label{SPqsnr_def}
\mathcal{Q} _{SP}\left(\lambda\right) \equiv \lambda^2\mathcal{H}\left(\lambda\right)
\end{equation}
where we have here emphasized that the QSNR in question is the one corresponding to single-parameter estimation through the ‘SP’ subscript. For the multiparameter setting we consider a generalization of Eq.~\eqref{SPqsnr_def}, 
\begin{equation}\label{MPqsnr_def}
    \mathcal{Q}_{MP}(\lambda_i,\lambda_j) \equiv \frac{|\lambda_i\lambda_j|}{[\boldsymbol{\mathcal{H}}^{-1}]_{\lambda_i,\lambda_j}} \ge \frac{|\lambda_i\lambda_j|}{\text{Cov}(\lambda_i,\lambda_j)}
\end{equation}
where the bounds follow from the element-wise manipulation of the quantum CRB in Eq.~\eqref{eq:MPQCRB}. The maximum possible quantum signal to noise ratio $\lambda_i^2/\text{Var}(\lambda_i)$ for the estimation of parameter $\lambda_i$ in a system with multiple unknown quantities is given by $\mathcal{Q}_{MP}(\lambda_i,\lambda_i)$, which is a strictly non-negative quantity. The off-diagonal components $\mathcal{Q}_{MP}(\lambda_i,\lambda_j)$ with $\lambda_i \ne \lambda_j$ relate to covariances and therefore can be negative when the measurement outcomes of $\lambda_i$ and $\lambda_j$ are negatively correlated. Note that each element of the multiparameter QSNR defined in Eq.~(\ref{MPqsnr_def}) is proportional to the determinant of the QFIM, ${\rm det}(\boldsymbol{\mathcal{H}})$, due to the matrix inverse operation appearing in Eq.~(\ref{eq:MPQCRB}). As such the QSNR vanishes when the QFIM is \emph{singular} since then ${\rm det}(\boldsymbol{\mathcal{H}})=0$. We discuss how to interpret and deal with a singular QFIM in Sec.~\ref{Control_Field}.

Let us explicitly consider the estimation of two arbitrary parameters, $\vec{\lambda} = \left(\lambda_A, \lambda_B \right)^T$. Equation~\eqref{eq:MPQCRB} requires the inverse of the $2\times 2$ QFIM,
\begin{equation}
    \label{Eq:Hinv}
\boldsymbol{\mathcal{H}}^{-1} = \frac{1}{{\rm det}(\boldsymbol{\mathcal{H}})} \begin{pmatrix}
		\boldsymbol{\mathcal{H}}_{\lambda_B,\lambda_B} & -\boldsymbol{\mathcal{H}}_{\lambda_A,\lambda_B} \\
		-\boldsymbol{\mathcal{H}}_{\lambda_A,\lambda_B} & \boldsymbol{\mathcal{H}}_{\lambda_A,\lambda_A}
	\end{pmatrix}
\end{equation}
where ${\rm det}(\boldsymbol{\mathcal{H}})=\boldsymbol{\mathcal{H}}_{\lambda_A,\lambda_A}\boldsymbol{\mathcal{H}}_{\lambda_B,\lambda_B}-\boldsymbol{\mathcal{H}}_{\lambda_A,\lambda_B}^2$. Thus we find,
\begin{equation}
    \label{Eq:MP_Thermo}	\mathcal{Q}_{MP}\left(\lambda_A,\lambda_A\right) = \lambda_A^2~ \frac{{\rm det}(\boldsymbol{\mathcal{H}})}{\boldsymbol{\mathcal{H}}_{\lambda_B,\lambda_B}}=\mathcal{Q}_{SP}(\lambda_A)-\frac{(\lambda_A\lambda_B\boldsymbol{\mathcal{H}}_{\lambda_A,\lambda_B})^2}{\mathcal{Q}_{SP}(\lambda_B)}~~
\end{equation} 
and similarly for $\mathcal{Q}_{MP}\left(\lambda_B,\lambda_B\right)$. For the off-diagonal terms,
\begin{equation}
    \label{Eq:MP_TK}	\mathcal{Q}_{MP}\left(\lambda_A,\lambda_B\right) = -\abs{\lambda_A\times \lambda_B}~\frac{{\rm det}(\boldsymbol{\mathcal{H}})}{\boldsymbol{\mathcal{H}}_{\lambda_A,\lambda_B}}
\end{equation} 
 These expressions immediately provide insight into the multiparameter estimation problem. First, we see that the multiparameter QSNRs $\mathcal{Q}_{MP}(\lambda,\lambda)$ can be decomposed into a piece corresponding to the single-parameter estimation of the same parameter $\mathcal{Q}_{SP}(\lambda)$ and a correction. This correction always \textit{lowers} the multiparameter QSNR relative to its single-parameter counterpart. The multiparameter QSNR is therefore upper-bounded by the corresponding single-parameter QSNR -- no additional precision in the estimation of parameter $\lambda$ may be obtained by uncertainties in other parameters, see also Appendix~\ref{Appendix:Uncertainties_multiparameter}. The magnitude of this correction is increased by the cross-correlation part of the QFIM $\boldsymbol{\mathcal{H}}_{\lambda_A,\lambda_B}$ but it is \emph{decreased} when the single-parameter QFI of the \emph{other} parameter is larger. This makes physical sense, because the multiparameter QSNR for parameter $\lambda$ should approach its single-parameter QSNR value when $\lambda' \ne \lambda$ is known with certainty (whereby the QFI for $\lambda'$ diverges and the deleterious correction vanishes).

This can be seen clearly by rearranging the equations to obtain the following identity: 
\begin{equation}
\frac{\mathcal{Q}_{MP}\left(\lambda_A,\lambda_A\right)}{\mathcal{Q}_{SP}(\lambda_A)} = \frac{\mathcal{Q}_{MP}\left(\lambda_B,\lambda_B\right)}{\mathcal{Q}_{SP}(\lambda_B)} =1-{\rm Cor}(\lambda_A,\lambda_B)^2
\end{equation}
where we have defined ${\rm Cor}(\lambda_A,\lambda_B)\!=\!\boldsymbol{\mathcal{H}}_{\lambda_A,\lambda_B}/\sqrt{\boldsymbol{\mathcal{H}}_{\lambda_A,\lambda_A}\boldsymbol{\mathcal{H}}_{\lambda_B,\lambda_B}}$ which is equal to the correlation ${\rm Cov}(\lambda_A,\lambda_B)/\sqrt{{\rm Var}(\lambda_A){\rm Var}(\lambda_B)}$ between measurements of $\lambda_A$ and $\lambda_B$ in the `best case scenario' when the quantum multiparameter CRB in Eq.~\eqref{eq:MPQCRB} is saturated.
These relations embody the fact that the relative degradation in measurement precision when there is uncertainty in both parameters applies equally to both parameters. The degradation is controlled by the \textit{correlation} between the parameters. As a consequence, in scenarios where the QFIM is singular we have ${\rm Cor}(\lambda_A,\lambda_B)\to 1$ and $\mathcal{Q}_{MP}\to 0$.

These results have implications for critical quantum metrology, which typically assumes that only a single parameter is to be estimated, and we have perfect knowledge of all other system parameters. However, Eq.~\eqref{Eq:MP_Thermo} for the two-parameter estimation scenario shows that uncertainty in one parameter can dramatically affect the sensing capability for another parameter. Therefore, quantum critical systems are not inherently good for parameter estimation unless the single-parameter QSNRs for \textit{all} unknown parameters are large.

In general, elements of the QFIM may depend explicitly on both the state's eigenvalues and eigenvectors~\cite{Paris_Quantum_Estimation}.
We now comment on a special but important case where the eigenvectors of the probe reduced density matrix do not explicitly depend on the parameters $\vec{\lambda}$ to be estimated. Then,
\begin{equation}
\label{eq:QFI_MP}	\boldsymbol{\mathcal{H}}_{\lambda_i,\lambda_j}  = \sum_k\frac{\partial_{\lambda_i} \rho_k(\vec{\lambda}) \times \partial_{\lambda_j} \rho_k (\vec{\lambda})}{\rho_k(\vec{\lambda})}\;,
\end{equation}
where $\rho_k(\vec{\lambda})$ are the parameter-imprinted eigenvalues of the probe reduced density matrix.

Furthermore, we note that if the probe's populations 
 are determined solely by a single observable $\Omega$, i.e. $\varrho_k\equiv \varrho_k(\Omega)$, then the QFIM elements follow as
\begin{equation}\label{eq:qfim_factor_gen}    \boldsymbol{\mathcal{H}}_{\lambda_i,\lambda_j} =  \mathcal{H}(\Omega)\times (\partial_{\lambda_i}\Omega ~ \partial_{\lambda_j}\Omega)
\end{equation}
where $\mathcal{H}(\Omega)=\sum_k (\partial_{\Omega} \varrho_k)^2/\varrho_k$ is an effective single-parameter estimation QFI. The factorized form of Eq.~\eqref{eq:qfim_factor_gen} immediately implies that the QFIM is singular, with $\text{det}(\boldsymbol{\mathcal{H}})=0$. Thus, the multiparameter QSNRs in Eq.~\eqref{MPqsnr_def} identically vanish, $\mathcal{Q}_{MP}(\lambda_i,\lambda_j)=0$. This tells us that under such a scenario, absolutely no information can be extracted about multiple unknown parameters from the measurements of the single observable $\Omega$. 
By contrast, if only $\lambda_i$ is to be estimated and all other parameters are known, then the corresponding single-parameter QSNR $\mathcal{Q}_{SP}(\lambda_i)$ is finite because the diagonal element of the QFIM $\boldsymbol{\mathcal{H}}_{\lambda_i,\lambda_i}$ is finite. As soon as we have two unknown parameters, $\lambda_i$ and $\lambda_j$, nothing can be said about either of them from measurements on the probe because the other QFIM elements come into play.

\begin{figure}[t]
	\centering
	\includegraphics[width=1.02\linewidth]{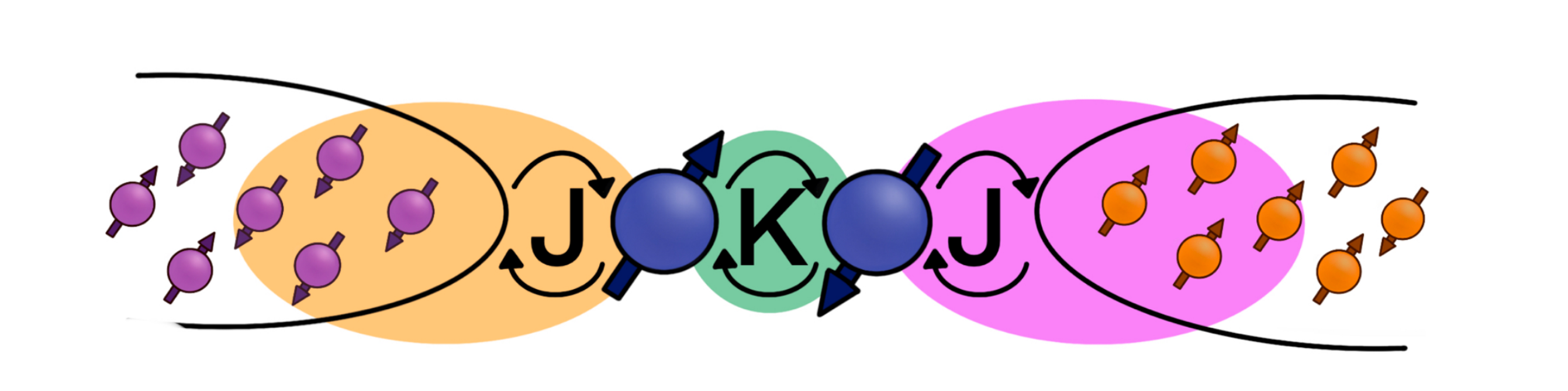}
\caption{Illustration of the 2IK model studied in this work. Two spin-$\tfrac{1}{2}$ `impurity' qubits comprising the probe are exchanged-coupled together, and each is also coupled to its own fermionic environment. We treat the fermionic environments explicitly as metallic leads involving a continuum of electronic states, appropriate to a realization of the model in a quantum nanoelectronics device. The full system allows nontrivial correlations and quantum entanglement to build up between the impurity probes and the fermionic environment of the leads. In particular, many-body physics associated with the probe-lead coupling $J$ favouring the Kondo effect, competes with the intra-probe coupling $K$ which favours local spin-singlet formation. This frustration produces a quantum critical point with macroscopic probe-environment entanglement and  fractionalized excitations.} 
	\label{Fig:Depiction}
\end{figure}


\section{Physical System and Model}\label{sec3:model}
We consider a simple probe system $\hat{H}_{\rm probe}=K~\hat{\vec{\mathbf{S}}}_{IL}\cdot\hat{\vec{\mathbf{S}}}_{IR}$ 
involving two coupled quantum spin-$\tfrac{1}{2}$ `impurity' qubits, where here and throughout we assume units such that $\hbar = 1$. The unique spin-singlet state of the isolated probe is the ground state when the exchange coupling between the impurities is antiferromagnetic $K>0$, whereas the degenerate spin-triplet state is the ground state for ferromagnetic coupling $K<0$.

Importantly, in our setup we model explicitly the environment to be probed by the impurities. We take the environment to be a fermionic bath comprising a continuum of electronic states, in the thermodynamic limit, which we divide into $\alpha\!=\!L$ and $\alpha\!=\!R$ regions (leads). Each probe impurity $\alpha$ is then coupled to its own bath $\alpha$ by an exchange coupling $J$, see Fig.~\ref{Fig:Depiction}. The full `two impurity Kondo' (2IK) model \cite{jayaprakash1981two,jones1988low,affleck1992exact,affleck1995conformal,mitchell2012two,sela2011exact,andrew_entropy_exact} reads,
\begin{equation}
	\label{Eq:Full_System_Hamiltonian_No_Field}
	\hat{H} = \Hat{H}_{E} + \hat{H}_{\rm probe} + J~\left(\hat{\vec{\mathbf{S}}}_{IL}\cdot\hat{\vec{\mathbf{s}}}_{EL} + \hat{\vec{\mathbf{S}}}_{IR}\cdot\hat{\vec{\mathbf{s}}}_{ER}\right)
\end{equation}
where the fermionic environment is described by,
\begin{equation}
	\label{Eq:Environment_Diag_Ham}
	\hat{H}_{E}= \sum_{\alpha=L,R} \hat{H}_{E;\alpha} = \sum_{\alpha=L,R} \sum_{k,\sigma}\epsilon_k\hat{c}^\dagger_{\alpha k\sigma}\hat{c}^{\phantom{\dagger}}_{\alpha k\sigma}
\end{equation}
where $\hat{c}_{\alpha k\sigma}^{(\dagger)}$ annihilates (creates) an electron in the single-particle momentum state $k$ with spin $\sigma=\uparrow$ or $\downarrow$ in bath $\alpha$. Here $\epsilon_k$ is the dispersion, which for simplicity we take to be linear at low energies, giving a standard flat (metallic) electronic density of states within a band of halfwidth $D$. The operator $\hat{\vec{\mathbf{s}}}_{E\alpha}=\tfrac{1}{2} \sum_{\nu\nu'} \hat{c}_{\alpha 0 \sigma}^{\dagger} \vec{\boldsymbol{\sigma}}_{\sigma\sigma'} \hat{c}^{\phantom{\dagger}}_{\alpha 0 \sigma'}$ describes the spin density of bath $\alpha$ at the probe position, where $\hat{c}_{\alpha 0 \sigma}$ is the corresponding \textit{local} bath orbital to which impurity probe $\alpha$ couples. 

We note that the 2IK model is \textit{gapless} when the leads are in the thermodynamic limit.


\subsection{Phases and critical point}
We consider the thermalized probe-lead system at equilibrium, treating the full many-body system exactly: non-perturbative renormalization effects at low temperatures produce macroscopic probe-lead entanglement through the Kondo effect~\cite{hewson1997kondo,jones1988low,bayat2012entanglement,bayat2014order}, and the backaction effect of the probe on the fermionic environment cannot neglected when considering the probe response. The 2IK model embodies a nontrivial competition between the frustrated magnetic interactions $K$ and $J$~\cite{jones1988low}. For $K/J \to 0$ we have two decoupled single-impurity Kondo models. 
For antiferromagnetic $J>0$ the Kondo effect produces strong-coupling physics at low temperatures $T\ll T_K$, with $T_K\sim D e^{-2D/J}$ a low-energy scale called the Kondo temperature. The impurity spin is dynamically screened below $T_K$ by surrounding conduction electrons in the environment by the formation of a many-body entanglement cloud \cite{bayat2010negativity}. On the other hand, for $J/K \to 0$, the leads are effectively decoupled and the full model reduces to just $\hat{H}_{\rm probe}$. The singlet and triplet states of the coupled impurities are essentially unaffected by the leads in this limit. 

Frustration between incompatible singlet-formation mechanisms produces a second-order quantum phase transition in the thermodynamic limit of the 2IK model. A nontrivial critical point~\cite{affleck1995conformal,mitchell2012two,andrew_entropy_exact} arises when the binding energy of the Kondo effect $T_K$ driven by $J$ matches the magnetic binding energy between the impurities $K$. The critical point at $K=K_c \sim T_K$ separates a Kondo phase for $K<K_c$ from a magnetic (RKKY) phase for $K>K_c$. At zero temperature, tuning $K$ through the critical point at $K_c$ results in a change in the many-body ground state of the system. However, signatures of criticality are observed over a window of $K$ around $K_c$ at finite temperatures.


\subsection{Impurity Quantum Metrology}
We now explore the different regimes of the 2IK model in the context of single- and multiparameter metrology, focusing on the estimation of $T$ and $K$.  The 2IK model captures a special case where the eigenvectors of the probe reduced density matrix do not explicitly depend on either $T$ or $K$, and therefore we may use Eq.~\ref{eq:QFI_MP} with $\vec{\lambda}= \left(T,K\right)^T$ to calculate the QFIM. As shown below, we find that the QFIM is singular for the 2IK model, and therefore estimation of the parameters $T$ and/or $K$ by measurements on the probe is impossible when there is uncertainty in both $T$ and $K$. We therefore first consider the unproblematic single parameter estimation scenario in the following sections. Then in Sec.~\ref{Sec:MultiParam_Background} we show that the QFIM singularity can be removed by applying a \emph{known} control field, thereby allowing us to recover true multiparameter estimation sensitivity in this system.

The global SU$(2)$ spin symmetry of the full system is preserved on the level of the probe reduced density matrix (RDM) obtained by tracing out the electronic leads. This allows us to construct the probe RDM in the spin eigenbasis, which for the two-impurity system is diagonal:
\begin{equation}
    \label{eq:No-B-RDM-form}
	\hat{\varrho}_{\rm probe} = \text{diag}\left(\varrho_{S}, \varrho_{{T;+1}}, \varrho_{{T;0}}, \varrho_{{T;-1}} \right),
\end{equation}
where $\varrho_S$ is the population of the reduced state spin-singlet, and $\varrho_{T;-1}=\varrho_{T;0}=\varrho_{T;+1}\equiv \varrho_T$ are the populations of the reduced state components of the spin-triplet (which are equal by symmetry when no external field acts).

Interestingly, even for the lead-coupled, many-body system, we can fully obtain $\hat{\varrho}_{\rm probe}$ from a single physical observable that can be measured on the probe system. We define the impurity spin-spin correlator (equivalent to the probe singlet fraction) as,
\begin{equation}
    \label{eq:Spin-Spin-Corr-Def}
	\mathcal{C} = \langle \hat{\vec{\mathbf{S}}}_{IL}\cdot\hat{\vec{\mathbf{S}}}_{IR} \rangle \equiv \Tr{\left (\hat{\vec{S}}_{IL}\cdot\hat{\vec{S}}_{IR}\right )~\hat{\varrho}_{\rm probe}} \;,
\end{equation}
from which it follows that $\mathcal{C}=\tfrac{3}{4}\left( \varrho_T-\varrho_S\right)$. Together with the normalization condition $\varrho_S+3\varrho_T=1$ we can determine all of the RDM elements in terms of the observable correlator $\mathcal{C}$ as:
\begin{eqnarray}
	\varrho_{S} &=& \frac{1}{4} - \mathcal{C} \;,   \label{eq:No-B-PES-Sol}\\
	\varrho_{T} &=& \frac{1}{4} + \frac{1}{3}\mathcal{C} \;.   \label{eq:No-B-PET-Sol}
\end{eqnarray}

We are now in a position to calculate the sensitivity in terms of the QFI via the correlator $\mathcal{C}$. Since the probe RDM is diagonal in the spin basis, the RDM eigenvectors are independent of the model parameters. For single-parameter estimation, where we assume there is only a single unknown parameter, Eq.~\eqref{eq:QFI_MP} reduces from a matrix to a scalar given by 
\begin{equation}
    \label{eq:QFI-SP-C}
	\mathcal{H}\left(\lambda\right) = \sum_i \frac{|\partial_\lambda \varrho_i|^2}{\varrho_i} =  \frac{|\partial_\lambda \mathcal{C}|^2}{\left(\tfrac{1}{4} - \mathcal{C}\right)\times\left(\tfrac{3}{4} + \mathcal{C}\right)}
\end{equation}
for $\lambda = T$ or $K$. Since the correlator $\mathcal{C}$ determines completely the probe RDM, it is equivalent to the SLD for this system and therefore corresponds to the \textit{ideal} measurement for metrological purposes, saturating the quantum CRB. 


\section{Large $K$-limit}\label{Sec:large_K}
We will first examine the single parameter estimation scenario for both temperature $T$ and coupling $K$. For simplicity and to provide physical insight into the behaviour of the full system, we focus here on an analytically tractable limiting case: the large-$K$ limit ($K/J\gg 1)$, where the fermionic leads play essentially no role in determining the reduced states of the probe. We may therefore approximate the full model as just that of the probe,
\begin{equation}
	\label{Eq:Large_K_Lim_Ham}
	\hat{H}_{KL} = \hat{H}_{\rm probe} \equiv K~\hat{\vec{\mathbf{S}}}_{IL}\cdot\hat{\vec{\mathbf{S}}}_{IR} \;.
\end{equation}
The Hamiltonian is readily diagonalized and the probe density matrix follows immediately: $\varrho_{{i}} =e^{-E_i/T}/ \mathcal{Z}$ are the thermal populations, properly normalised by the partition function $\mathcal{Z}=\sum_i e^{-E_i/T}$. Here $E_S=-3K/4$ is the energy of the two-impurity probe spin-singlet state, and $E_{T;S^z}=+K/4$ is the energy of the three degenerate components of the spin-triplet state. Importantly, all populations are a function of the single rescaled parameter $y\!=\!K/T$. They can also be obtained in terms of the correlator $\mathcal{C}$.

\begin{figure*}
	\begin{center}
		\includegraphics[width=1\linewidth]{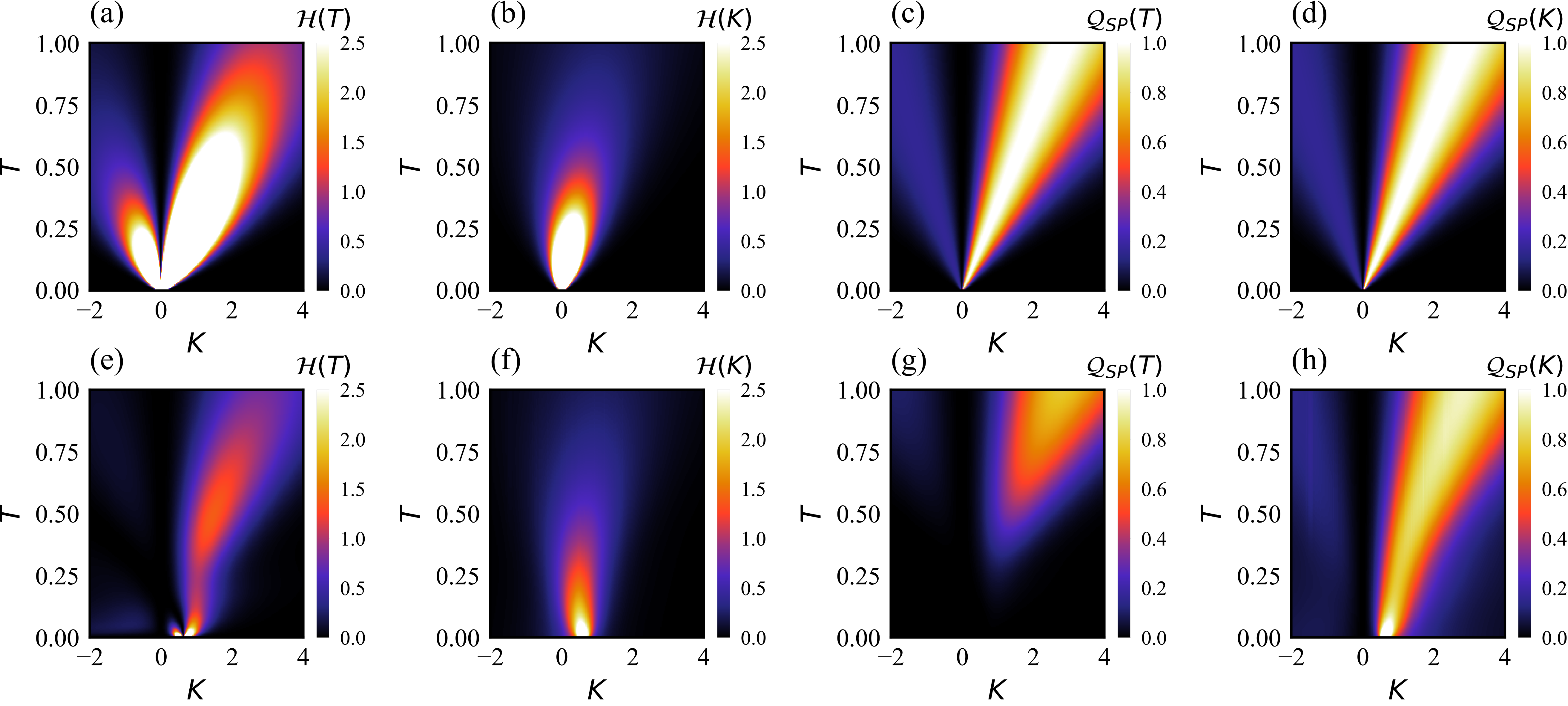}
	\end{center}
	\caption{\textbf{Single-parameter estimation of $T$ and $K$ in the 2IK model.} \emph{Top row (panels a-d):} Results in the large-$K$ limit extracted from analytical solution. \emph{Bottom row (panels e-h):} Numerical results for the full 2IK model from NRG. Columns 1 and 2 show the QFI for thermometry $\mathcal{H}(T)$ and for estimation of the probe coupling strength $\mathcal{H}(K)$ as a function of $T$ and $K$, whereas columns 3 and 4 show the corresponding single-parameter QSNR responses $\mathcal{Q}_{SP}\left(T\right)$
and $\mathcal{Q}_{SP}\left(K\right)$.}
	\label{Fig:KLIM_NoB_Thermo}
\end{figure*}

Although we consider only the probe Hamiltonian explicitly here, we take the fermionic leads implicitly and assume thermalization has occurred. The reduced states of the probe are therefore equivalent to the isolated probe states, and we can use the exact populations $\varrho_i$ obtained in Eqs.~\eqref{eq:No-B-PES-Sol} and \eqref{eq:No-B-PET-Sol} to calculate the single-parameter QFI for estimation for $T$ or $K$ by using Eq.~\eqref{eq:QFI-SP-C}. 
In this limit we find for thermometry,
\begin{equation}
	\label{Eq:KLIM_NoB_Thermo_QFI}
	\mathcal{H}\left(T\right) = \frac{3~e^{K/T}~K^2}{\left(3 + e^{K/T}\right)^2 T^4} \;,
\end{equation}
and for estimation of $K$,
\begin{equation}
	\label{Eq:KLIM_NoB_Kometry_QFI}
	\mathcal{H}\left(K\right) = \frac{3~e^{K/T}}{\left(3 + e^{K/T}\right)^2 T^2} \;.
\end{equation}
We show the behavior of these quantities in Fig.~\ref{Fig:KLIM_NoB_Thermo}(a,b). 

Evidently, the precision of parameter estimation is highly dependent on the singlet-triplet transitions in this limit. The best performance for estimating a given parameter by making local measurements on the probe is obtained when the population transfer between singlet and triplet states upon changing that parameter is maximal. Since the probe singlet and triplet states are separated by a single energy gap $|K|$, this naturally happens when $T \sim |K|$.

A second interesting feature of the single-parameter QFIs obtained in the large-$K$ limit is the role of the \textit{sign} of $K$. This is because the degeneracy of the ground and excited probe states gets swapped when the sign of $K$ is flipped. Specifically, for positive (antiferromagnetic) coupling $K>0$, the ground state is the unique spin-singlet state  $\ket{S} = \tfrac{1}{\sqrt{2}}\left(\ket{\uparrow\downarrow} - \ket{\downarrow\uparrow}\right)$, while the excited states are the three degenerate components of the spin-triplet, $\ket{T;+1}=\ket{\uparrow\uparrow}$, $\ket{T;0} = \tfrac{1}{\sqrt{2}}\left(\ket{\uparrow\downarrow} + \ket{\downarrow\uparrow}\right)$ and $\ket{T;-1}=\ket{\downarrow\downarrow}$. For negative (ferromagnetic) coupling $K<0$, it is the ground state that is degenerate and the excited state that is unique. As explored in previous works~\cite{correa2015individual, Campbell_Percision_Thermo_QSL}, systems with excited state degeneracies are known to give higher thermometric precision. We observe the same phenomenon here: better performance is obtained for $K>0$ than $K<0$. This gives rise to the asymmetric structure of the QFI plots in both Fig.~\ref{Fig:KLIM_NoB_Thermo}(a) and (b) around $K=0$. We note that the largest QFI arises at \textit{low} temperatures and coupling strengths. In particular, it might seem counterintuitive that the best sensitivity to coupling strength $K$ is obtained when the probe impurities are actually decoupled, $K=0$. This illustrates the need for the QSNR rather than the QFI itself when interpreting metrological capability. We further note that $\mathcal{H}(T)$ and $\mathcal{H}(K)$ look very different, even though the underlying probe populations in the large-$K$ limit depend only on the single rescaled parameter $y\!=\!K/T$, and so we expect the corresponding QFIs for $K$ and $T$ to be simply related. Again, the QSNR helps to uncover these similarities.

First, we note that from Eq.~\eqref{eq:QFI-SP-C} we may write $\mathcal{H}(\lambda)=\mathcal{H}(y) / |\partial_\lambda y|^2$. Since $\partial_K y=1/T$ and $\partial_T y=-K/T^2$, it follows immediately that $\mathcal{H}(T)\times T^2 = \mathcal{H}(K)\times K^2$. This is precisely the definition of the QSNR in Eq.~\eqref{SPqsnr_def}, such that
\begin{equation}
	\label{Eq:KLIM_QSNR_Equiv}
	\mathcal{Q}_{SP}\left(T\right) = \mathcal{Q}_{SP}\left(K\right) = \frac{3~e^{K/T}~K^2}{\left(3 + e^{K/T}\right)^2 T^2} \;,
\end{equation}  
which is a universal function of the single parameter $y=K/T$.
The QSNR results for the 2IK system in the large-$K$ limit are presented in Fig.~\ref{Fig:KLIM_NoB_Thermo}(c,d) and demonstrate clearly the physical features discussed above. In particular, we see that the QSNR is indeed identical for single-parameter $T$ and $K$ estimation, with the maximum sensitivity attained along the line $K\simeq 2.85 T$ for $K>0$ with $\mathcal{Q}_{Max}\simeq 1$, whereas for $K<0$ the maximum sensitivity is lower with  $\mathcal{Q}_{Max}\simeq 1/6$ along the line $K\simeq -2.16T$. We attribute the boosted robustness to measurement noise in the antiferromagnetic regime to the enhanced probe degeneracy for the excited state in this case.

Although the large-$K$ limit is over-simplified, we expect certain qualitative features (such as the difference between positive and negative $K$) to carry over to the full solution. In particular, we remind that the QFI and QSNR signatures of the 2IK probe are fully determined by a single measureable observable, the spin-spin correlator $\mathcal{C}$, not only in the large-$K$ limit but also in the full lead-coupled model. Although many-body effects conspire to produce richer physics in the full lead-coupled system that are naturally reflected in a more complex structure for $\mathcal{C}$, we may still use Eq.~\eqref{eq:QFI-SP-C} to extract metrological properties. On the other hand, we do not generally expect $\mathcal{Q}_{SP}(T)=\mathcal{Q}_{SP}(K)$ when the leads are attached because the probe RDM eigenvalues are then no longer simple functions of the single rescaled parameter $K/T$ and competition with other scales ($J,D,T_K$) will become important. We refer to Appendix~\ref{Appendix:Narrow_Band_Limit} where the nontrivial competition that arises when we have competing energy scales can already been seen in the narrow band limit (i.e. when $J/D \gg 1$ and wherein the electronic states in the leads can be approximated by a single local orbital in real space). 


\section{2IK model: NRG results}\label{Sec:NRG_Results}
When a full continuum of electronic states is included in the metallic leads (which constitute the environment), the 2IK model, Eq.~\eqref{Eq:Full_System_Hamiltonian_No_Field}, is a famous strongly-correlated many-body problem \cite{jones1988low,affleck1995conformal,andrew_entropy_exact} whose solution requires the use of sophisticated methods. Here we use Wilson's NRG method \cite{bulla2008numerical,weichselbaum2007sum,mitchell2014generalized,*stadler2016interleaved} to obtain the spin-spin correlation function $\mathcal{C}=\langle \hat{\vec{\mathbf{S}}}_{IL}\cdot\hat{\vec{\mathbf{S}}}_{IR} \rangle$ numerically as a function of temperature $T$ and couplings $K$ and $J$. We set the conduction electron bandwidth $D=1$, use NRG discretization parameter $\Lambda=2.5$ and keep $N_s=8000$ states at each step of the iterative diagonalization procedure. As previously emphasized, a knowledge of $\mathcal{C}$ (and its derivatives) is sufficient to determine fully the single parameter metrological capability of the 2IK probe -- and so NRG, which provides numerically-exact access to this quantity, is an ideal tool. The evolution of $\mathcal{C}$ as a function of $T$ and $K$ is smooth in the full model, although at $T=0$ we see a discontinuity in $\partial_K \mathcal{C}$ at $K=K_c$, indicating the existence of a second-order quantum phase transition~\cite{affleck1995conformal}.


\subsection{Overview of metrological phase diagram}
Fig.~\ref{Fig:KLIM_NoB_Thermo}(e-h) show the NRG results for the full lead-coupled 2IK model, as a function of $K$ and $T$. We set $J=1$ here, and find that the critical point is located at $K=K_c\approx 0.62$. Several qualitative features are reminiscent of the large-$K$ limit results in panels (a-d). In particular, the thermometric QFI $\mathcal{H}(T)$ shows a split two-lobe structure; but in the full model this behavior is pushed to low temperature $T\ll J$ and is centred at the critical point $K=K_c$ rather than $K=0$. Likewise for $\mathcal{H}(K)$ we see a single intense flair, but again it is now located at $K=K_c$.

The QSNR results in Fig.~\ref{Fig:KLIM_NoB_Thermo}(g,h) tell the clearest story, however. Despite the large thermometric QFI near the critical point, this enhanced precision arises only at low temperatures where the thermometric signal is also small. Overall the thermometric QSNR is surprisingly poor at low temperatures. Only at larger $T$ and $K$ do we see good temperature estimation capability due to the finite $J$ (or $T_K$) scale. This is because the measured probe observable $\mathcal{C}$ does not change appreciably with temperature when $T\ll T_K$, even at the critical point $K=K_c$.  On the other hand, the QSNR sensitivity to single-parameter estimation of the coupling constant $K$ is strongly enhanced near the critical point, especially at low temperatures. This is due to the 
the sharp crossover in $\mathcal{C}$ as $K$ is tuned in the vicinity of the critical point at $K_c$.

We argue that these are generic features near a second-order quantum phase transition, expected on general grounds from a renormalization group (RG) perspective. In the critical regime, the physics of the system is controlled only by the critical fixed point at low temperatures, with the temperature-dependence of physical observables scaling weakly with RG irrelevant, or at best marginal, operators  \cite{sachdev1999quantum}. By contrast, the dependence on a parameter driving the transition will typically be strong, since by definition its scaling is controlled by RG relevant operators.


\subsection{Universal Kondo regime}
\begin{figure}[t]
	\centering
	\includegraphics[width=0.85\linewidth]{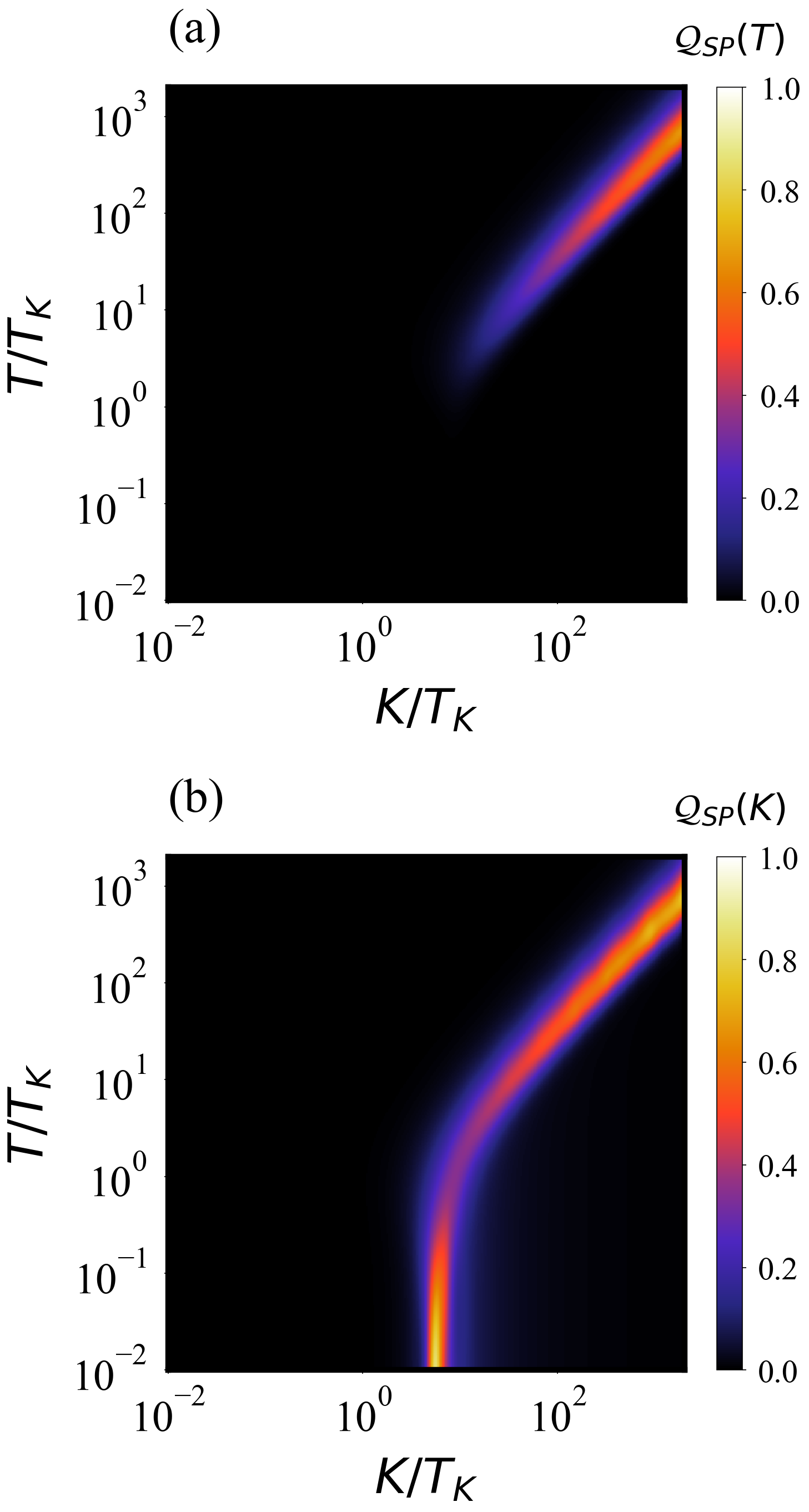}
\caption{\textbf{QSNR of the full 2IK model in the universal Kondo regime.} \emph{Top panel (a):} QSNR for thermometry $\mathcal{Q}_{SP}\left(T\right)$. \emph{Bottom panel (b):} QSNR for coupling constant $\mathcal{Q}_{SP}\left(K\right)$. NRG calculations performed using $J=0.15$, for which the Kondo temperature is found to be $T_K \approx 10^{-7}D$ and the critical point is at $K=K_c\approx 6 T_K$. Note: axes are rescaled in terms of $T_K$ and on a log scale.} 
	\label{Fig:Sample_Around_TK}
\end{figure}

The full 2IK model supports a nontrivial quantum phase transition that separates two distinct regimes \cite{jones1988low,affleck1995conformal,mitchell2012two}. For $K<K_c$ we have essentially two separated single-impurity Kondo models, whereas for $K>K_c$ the two impurity probes lock up into a local spin-singlet and effectively decouple from the leads. Close to the critical point $K\sim K_c$ quantum fluctuations develop on long length and time scales and both impurities and leads become strongly entangled in one composite \cite{bayat2012entanglement,bayat2014order}. The critical value of $K_c$ depends on $J$ and is on the order of the single-channel $T_K$, which therefore gets small very quickly. In this section we use NRG to investigate the universal Kondo regime around $T\sim T_K$ and $K\sim T_K$ for $J=0.15$, for which we find $T_K\approx 10^{-7}$ and $K_c\approx 6T_K$. The reason for this choice is that the nonuniversal physics associated with the bare scales $D$ and $J$ is then unimportant. The critical point at $K_c$ is determined from NRG calculations by tuning $K$ to achieve a vanishing Fermi liquid scale.

In Fig.~\ref{Fig:Sample_Around_TK} we consider the QSNR for single-parameter estimation of $T$ and $K$ in this regime, with results obtained by NRG. We see very clearly from the numerical results the onset of critical physics and the demarcation of the two phases of the model, when using the impurity probes for metrology -- especially so for $\mathcal{Q}_{SP}(K)$ in the lower panel.

For thermometry, Fig.~\ref{Fig:Sample_Around_TK}(a), the measurement sensitivity at low temperatures $T\sim T_K$ is almost entirely lost in the Kondo phase $K<K_c$. This is expected from the results for the single-impurity Kondo probe explored in Ref.~\cite{Mihailescu_Thermometry}, because only a small amount of information about the state is imprinted locally on the impurity probe due to the macroscopic Kondo entanglement with the leads. For $K>K_c$ however, we see that singlet-triplet transitions on the probe impurities give good measurement sensitivity when the temperature $T$ is on the order of the renormalized probe  gap, which scales as~$\sim K$.

In  Fig.~\ref{Fig:Sample_Around_TK}(b) we investigate  $\mathcal{Q}_{SP}(K)$ in the same system. In the Kondo phase $K<K_c$ we again expect low measurement sensitivity because the probe populations do not change much with $K$ when both probes are separately being Kondo screened by their attached leads. But for $K>K_c$ we again see enhanced sensitivity for $K\sim T$ due to the internal singlet-triplet transitions on the probe. The major difference is that $\mathcal{Q}_{SP}(K)$ is also very sensitive to the critical point in the model, with boosted precision for measurement of the coupling constant around $K_c$ at low temperatures. This makes intuitive sense since at low-$T$, changing $K$ away from its critical value of $K_c$ causes dramatic changes, driving the system into one or other of the stable phases of the model. 


\section{2IK model: exact results near criticality}\label{sec:exact_results}
We now examine carefully the full 2IK model in the close vicinity of the critical point. For a given $J$, the critical point at $K=K_c$ is characterized by a single energy scale $T_K$, equivalent to the usual single-impurity Kondo scale \cite{jones1988low}. When $T_K \ll J,D$, physical properties are universal scaling functions of the rescaled parameter $T/T_K$, controlled by the 2IK quantum critical fixed point \cite{affleck1995conformal}. However, detuning the impurity coupling introduces a finite relevant perturbation $\delta K=K-K_c$ which destabilizes the critical point and generates a nontrivial RG flow towards a Fermi liquid fixed point on the new scale of $T^*$ \cite{sela2011exact,andrew_entropy_exact}. For $T\ll T^*$ the system flows towards one of the two stable phases of the model, depending on the sign of the perturbation. For $\delta K<0$ the system flows to the Kondo phase, whereas for $\delta K>0$ the system flows to the local inter-impurity singlet phase. However, for small perturbations $|\delta K|\ll T_K$, we have good scale separation $T^*\ll T_K$, with $T^*$ given by \cite{affleck1995conformal}
\begin{equation}
    \label{Eq:T_star_Def}
	\frac{T^*}{T_K} = c~\left(\frac{K - K_C}{T_K}\right)^2 \;,
\end{equation}
where $c$ is a constant. 
For $T\ll T_K$ physical quantities are universal functions of the rescaled parameter $T/T^*$ and are entirely characteristic of the 2IK quantum critical point, independent of microscopic details. Remarkably, in this universal critical regime the 2IK admits an exact analytical solution \cite{sela2011exact,andrew_entropy_exact} -- despite it being a strongly-correlated many-body problem. In particular, the entropy flow along the crossover on the scale of $T^*$ is known analytically in closed form \cite{andrew_entropy_exact},
\begin{equation}\label{eq:entropy1}
	S\left(T\right) = \frac{1}{2}\log{2} + \bar{S}\left(\frac{T}{T^*}\right) \qquad\qquad :~T\ll T_K
\end{equation}
where $\bar{S}$ is defined as,
\begin{equation}\label{eq:entropy2}
	\bar{S}\left(t\right) = \frac{1}{t}\left[\psi\left(\frac{1}{2} + \frac{1}{t}\right) - 1 \right] - \log\left[\frac{1}{\sqrt{\pi}} \Gamma\left(\frac{1}{2} + \frac{1}{t}\right)\right]
\end{equation}
with $\Gamma(\cdot)$ and $\psi(\cdot)$ being the gamma and the digamma functions, respectively. Full NRG calculations for the 2IK entropy performed in the critical regime $T^*\ll T_K$ confirm Eqs.~\eqref{Eq:T_star_Def} and \eqref{eq:entropy1} precisely. Taking the standard operational definition of the Kondo temperature through $S(T=T_K)=\log(2)$ we extract the constant $c \simeq 0.035$ from the NRG thermodynamics.

For the purposes of single-parameter estimation, we need access to the correlator $\mathcal{C}=\langle \hat{\vec{\mathbf{S}}}_{IL}\cdot\hat{\vec{\mathbf{S}}}_{IR} \rangle$ so that we may compute the QFI via Eq.~\eqref{eq:QFI-SP-C}. The behavior of this correlator has not previously been discussed in the 2IK critical region. However, here we note that $\mathcal{C} = \partial_K~\mathcal{F}$
is an exact identity, where $\mathcal{F} = - T\ln{\mathcal{Z}}$ is the equilibrium thermodynamic free energy (grand potential), and $\mathcal{Z}$ is the partition function of the full system.
 Meanwhile, the thermodynamic entropy is also related to the free energy, $S=-\partial_T \mathcal{F}$. Therefore at thermal equilibrium we may apply a Maxwell relation to connect the entropy to the correlator,
\begin{equation}
	\partial_T\partial_K\mathcal{F} = \partial_K\partial_T\mathcal{F}
\end{equation}
Thus it follows that,
\begin{equation}
	\partial_T\mathcal{C} = - \partial_K S
\end{equation}
which holds as an exact identity for any $T$ and $K$ (not just in the critical region). 
However, the derivative on the right hand side can be evaluated explicitly using Eqs.~\eqref{eq:entropy1} and \eqref{eq:entropy2} if we confine our attention now to the critical region. This yields,
\begin{align}
    \label{eq:Exact_C}
\partial_T\mathcal{C}(T,K)=  \frac{2c~\delta K \left[T~T_K - c~\delta K^2~\psi'\left(\Phi\right)\right]}{T^2~T_K^2}
\end{align}
where we have defined $\Phi\equiv \Phi(T,K) = \tfrac{1}{2} + \tfrac{T^*}{T} = \tfrac{1}{2}+\tfrac{c~\delta K ^2}{T~T_K}$, and with $\psi'$ the trigamma function. An exact expression for the impurity spin-spin correlation function itself can now be obtained by integrating
\begin{align}
	\mathcal{C}\left(T,K\right) &= - \int dT~\partial_K S\left(T,K\right) \nonumber\\ &= \frac{2c~\delta K \left[\log{T} + \psi\left(\Phi\right)\right]}{T_K} + \mathcal{C}^*\label{eq:corr}
\end{align}
where the constant of integration $\mathcal{C}^*=\mathcal{C}(T=0,K=K_c)$ is found to be the value of the spin-spin correlator at the critical point. This is determined by noting that the indefinite integral is defined up to a function of $K$, and that $\mathcal{C}(T\ll T_K,K_C)=\mathcal{C}^*$ and $\mathcal{C}(T\to \infty,K)=0$. Although $\mathcal{C}^*$ depends in general on $J$, it can be calculated numerically with NRG. With this information, we have the full temperature and coupling dependence of the correlator in the critical region via Eq.~\eqref{eq:corr}. We note that similar Maxwell relations have been used recently in reverse, to determine the entropy changes for a process from measureable observables in experiments on quantum devices \cite{han2022fractional,child2022entropy}. 

We can now take the derivative with respect to $K$ to obtain,
\begin{equation}\label{eq:exact_c_k}
	\partial_K\mathcal{C}\left(T,K\right) = \frac{2c~\left[T~T_K\left(\log{T} + \psi\left(\Phi\right)\right) + 2c~\delta K^2~\psi'\left(\Phi\right)\right]}{T~T_K^2}
\end{equation}
We validate these analytic results in Appendix~\ref{Appendix:Validation_Exact_Result} by explicit comparison to NRG results in the universal regime.

\begin{figure}[t]
	\centering
	\includegraphics[width=1.02\linewidth]{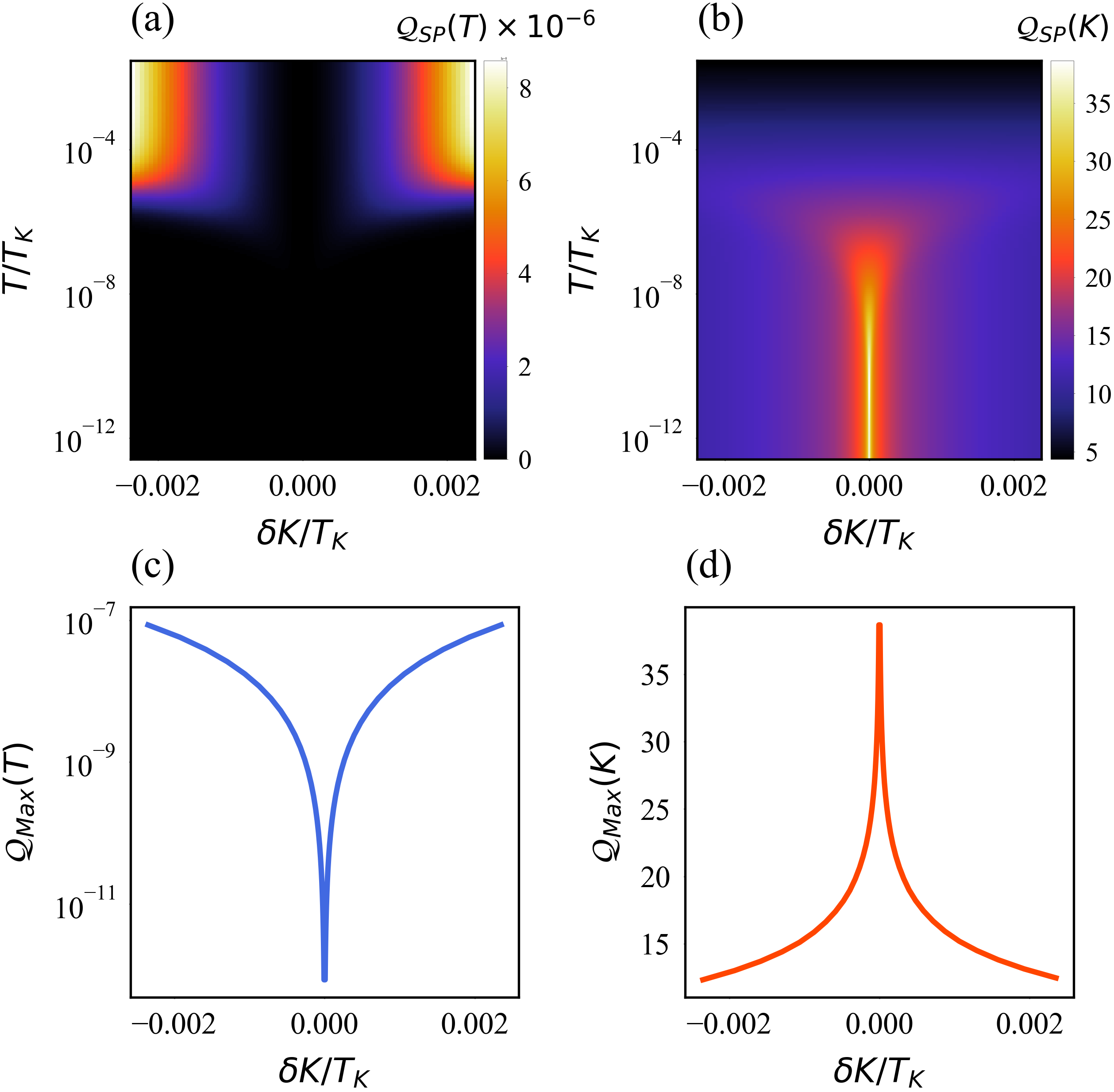}
	\caption{\textbf{Single parameter estimation of $T$ and $K$ in the universal quantum critical regime of the full 2IK model.} (a) Exact analytic result for the critical thermometric QSNR from Eq.~\eqref{eq:Exact_Thermo_KC} as a function of rescaled temperature $T/T_K$ and coupling detuning $\delta K/T_K$. (b) Corresponding QSNR for measuring the coupling constant $K$, obtained from  Eq.~\eqref{eq:Exact_Kermo_KC}. (c,d) Maximum QSNR in the critical regime for each value of the detuning $\delta K/T_K$. Results obtained for $J=1$. } 
	\label{fig:exactQSNR}
\end{figure}

We now have everything we need to compute the QFIs through Eq.~\eqref{eq:QFI-SP-C}. Our exact analytic expressions for the corresponding QSNRs in the critical region follow as,
\begin{equation}
    \label{eq:Exact_Thermo_KC}
	\mathcal{Q}_{SP}\left(T\right) = \frac{4c^2~\delta K^2\left[T~T_K - c~\delta K^2~\psi'\left(\Phi\right)  \right] ^2}{T^2~T_K^4~\left( \tfrac{1}{4} - \mathcal{C}\right)\times \left( \tfrac{3}{4} + \mathcal{C}\right)}
\end{equation}
and
\begin{equation}
    \label{eq:Exact_Kermo_KC}
	\mathcal{Q}_{SP}\left(K\right) = \frac{4c^2 K^2\left[T~T_K\left( \log{T} + \psi\left(\Phi\right) \right)+ 2c~\delta K^2~\psi'\left(\Phi\right)\right]^2}{T^2~T_K^4~\left( \tfrac{1}{4} - \mathcal{C}\right)\times \left( \tfrac{3}{4} + \mathcal{C}\right)}
\end{equation}
We present the corresponding exact QSNR results for the critical region in Fig.~\ref{fig:exactQSNR} using $J=1$, for which we find from NRG that $K_c\simeq 0.618$, $T_K\simeq 0.362$ and $\mathcal{C}^*\simeq -0.385$. We note that although Eqs.~\eqref{eq:Exact_Thermo_KC} and \eqref{eq:Exact_Kermo_KC} involve the exact correlator $\mathcal{C}$ from Eq.~\eqref{eq:corr}, in the universal regime of interest where $\delta K/T_K \lesssim 10^{-2}$ deviations of $\mathcal{C}$ away from its critical value $\mathcal{C}^*$ are tiny. Therefore our results are essentially indistinguishable if one replaces the functions $\mathcal{C}$ in the denominator of the expressions for the QSNR with the constant $\mathcal{C}^*$.
Our results show a highly nontrivial evolution of  measurement sensitivity in the quantum critical regime of this model. For thermometry, perhaps surprisingly, strong quantum critical correlations do not help with equilibrium measurement sensitivity at low temperatures. Physically, this is because in the region of parameter space described by the critical fixed point ($T^*\ll T \ll T_K$ with $T^*\sim \delta K^2$), the value of the probe spin-spin correlation function is $\mathcal{C}\simeq \mathcal{C}^*$ and has very little temperature dependence. Therefore $\partial_T \mathcal{C}$ is small and the corresponding QSNR is small. In particular, for $\delta K=0$ at the critical point, there is no RG flow for $T\ll T_K$ and $\delta_T \mathcal{C}\simeq 0$, meaning that we have zero measurement sensitivity to temperature at the critical point below the Kondo temperature. This is confirmed in Fig.~\ref{fig:exactQSNR}(a,c). This behavior is expected from the single-probe results of Ref.~\cite{Mihailescu_Thermometry}, where strong probe-environment entanglement hinders thermometric precision when measurements are only made on the probe. Likewise here, only a small amount of information about the environment temperature is imprinted on the probes when the probe-environment composite is in a strongly multipartite many-body entangled state. Indeed, in the small $T$ and small $\delta K$ limit we extract the asymptotic behavior from our exact solution 
$\mathcal{Q}_{SP}(T)\sim T^4\delta K^2/(a\delta K^8+T^4)$ with $a\simeq 0.04$, which is strongly suppressed near the critical point. Good thermometric precision is instead obtained at much higher temperatures outside the critical window, where the probes act essentially as coupled thermal qubits.

On the other hand, measuring other model parameters such as the internal probe coupling constant $K$ is a very different story. This is because changing $\delta K$ induces a change in the scale $T^*$ which pushes the system out of the critical window. The correlator $\mathcal{C}$ then changes significantly over a narrow range of $\delta K$ near the critical point. As such, the derivative $\partial_K \mathcal{C}$ is strongly increased at lower temperatures and smaller $\delta K$, where the critical window is narrower and small parameter changes have the largest effect. Indeed, as shown in Fig.~\ref{fig:exactQSNR}(b,d) measurement sensitivity is in fact \emph{diverging} rather than vanishing at the critical point, with 
$\mathcal{Q}_{SP}(K)\sim \log^2(bT+\delta K^2)$ with $b\simeq 0.13$ at small $T$ and $\delta K$. Our exact results in the close vicinity of the critical point are consistent with the broader NRG results for the QSNR presented in Figs.~\ref{Fig:KLIM_NoB_Thermo} and \ref{Fig:Sample_Around_TK}.  We note that the behavior discussed above should arise in any system in the Ising universality class \cite{affleck1995conformal} of boundary critical phenomena. 

Finally we comment on the role of fractionalization phenomena for metrology. Away from the critical point of the 2IK model, all of the probe degrees of freedom are quenched at low enough temperatures -- either by formation of a decoupled probe spin-singlet state for large $K\gg K_c$ or through the Kondo effect by environment-probe entanglement for $K\ll K_c$. Good measurement precision is then afforded by overcoming the excitation gap so that the reduced probe state populations become sensitive to changes in the parameter of interest. In the critical regime however, we have a somewhat different story. Here, the probe degrees of freedom are only partially screened: at the critical point, a degree of freedom remains unscreened even down to zero temperature, due to the frustration driving the phase transition \cite{affleck1995conformal}. In the critical regime,  it is the fate of this residual critical degree of freedom that is responsible for metrological performance when one manipulates the system by changing temperature or model parameters. Remarkably, this degree of freedom in the 2IK model is not just a qubit spin or electron, but a fractionalized \textit{Majorana} fermion \cite{affleck1995conformal,gan1995mapping,mitchell2012two,sela2011exact,andrew_entropy_exact}, with an exotic $\tfrac{1}{2}\ln(2)$ residual entropy -- see Eq.~\eqref{eq:entropy1}. This Majorana is localized on the probe, and we measure it when we make measurements on the probe near the critical point. Instead of using qubits or electronic degrees of freedom for sensing, here we effectively leverage the unusual properties of the Majorana fermion when we do metrology near the 2IK critical point.


\section{Multiparameter estimation} \label{Sec:MultiParam_Background}
 
We now consider multiparameter estimation in the 2IK model for $T$ and $K$. 
The QFIM follows in this case from Eq.~\eqref{eq:QFI_MP} with $\vec{\lambda}=(T,K)^T$. However, since the probe reduced state populations $\varrho_k \equiv \varrho_k(\mathcal{C})$ in Eq.~\eqref{eq:No-B-RDM-form} are entirely determined by the probe spin-spin correlator $\mathcal{C}$, then Eq.~\eqref{eq:qfim_factor_gen} holds and the QFIM is singular. The 2IK model is therefore a prime example where the single- and multiparameter estimation schemes yield totally different results. For example, for the estimation of the temperature $T$, any uncertainty in $K$ completely collapses the QSNR, and vice versa. Multiparametric QFIM singularities are therefore crucial to identify in any practical setup.

\subsection{Control field}\label{Control_Field}
Here we demonstrate that when the QFIM is singular, metrological sensitivity can be recovered by applying a {\it known} control field. In the 2IK model, the singularity in the QFIM hindering multiparameter estimation is a consequence of the large SU(2) spin symmetry of the coupled spin-$\tfrac{1}{2}$ qubit probes, which tightly constrains the properties of the probe reduced states. One might expect similar metrological problems in other systems with many conserved quantities (especially so for the class of integrable systems). For the 2IK model, breaking the SU(2) spin symmetry provides more flexibility and a route to precision parameter estimation. Here we focus on adding a known control field $B$ to our system, which we take to be just a simple Zeeman magnetic field oriented along the $z$-direction.
The model therefore becomes,
\begin{equation}
	\label{eq:2ik_B}
	\hat{H} = \Hat{H}_{E} + \hat{H}_{\rm probe} + J~\left(\hat{\vec{\mathbf{S}}}_{IL}\cdot\hat{\vec{\mathbf{s}}}_{EL} + \hat{\vec{\mathbf{S}}}_{IR}\cdot\hat{\vec{\mathbf{s}}}_{ER}\right)+\hat{H}_{\rm field}
\end{equation}
with control field Hamiltonian $\hat{H}_{\rm field}=B\hat{S}^z_{\rm tot}$, where the total spin projection is decomposed into $\alpha=L,R$ probe and environment parts $\hat{S}^z_{\rm tot}=
\sum_{\alpha} \left (\hat{S}_{I\alpha}^z +\hat{s}_{E\alpha;{\rm tot}}^z \right )$ with $\hat{S}^z_{I\alpha}$ the spin projection for impurity probe $\alpha$ and with $\hat{s}_{E\alpha;{\rm tot}}^z=\tfrac{1}{2}\sum_{k}\left( c_{\alpha k \uparrow}^{\dagger}c_{\alpha k \uparrow}^{\phantom{\dagger}}- c_{\alpha k \downarrow}^{\dagger}c_{\alpha k \downarrow}^{\phantom{\dagger}}\right )$ an operator for the total lead-$\alpha$ (environment) spin projection. We note that the results for a field applied locally to the probe rather than globally to the full probe-environment system are qualitatively and quantitatively very similar.

For $B\ne 0$ the full SU(2) spin symmetry is lifted, but a U(1) symmetry corresponding to conserved total $S^z_{\rm tot}$ remains. The probe reduced density matrix retains its diagonal structure in the spin basis Eq.~\eqref{eq:No-B-RDM-form}, but now the triplet populations are not equal, $\varrho_{T;-1}\ne \varrho_{T;0} \ne \varrho_{T;+1}$. In the full lead-coupled 2IK model, there is no simple exact relation between these reduced state populations. 
The probe spin-spin correlator $\mathcal{C}\!=\!\langle \hat{\vec{\mathbf{S}}}_{IL} \cdot \hat{\vec{\mathbf{S}}}_{IR} \rangle\!\equiv\! \Tr{\hat{\vec{\mathbf{S}}}_{IL} \cdot \hat{\vec{\mathbf{S}}}_{IR} ~\hat{\varrho}_{\rm probe}}$ is no longer sufficient to fully determine the probe reduced state. We therefore introduce two other physically-motivated and experimentally-feasible observables to fully characterize the probe: the probe magnetization $\mathcal{M}=\langle \hat{S}^z_{IL}+\hat{S}^z_{IR}\rangle\equiv \Tr{\left(\hat{S}_{IL}^z +  \hat{S}_{IR}^z\right)~\hat{\varrho}_{\rm probe}}$ and $\mathcal{\chi}=\langle (\hat{S}^z_{IL}+\hat{S}^z_{IR})^2\rangle\equiv \Tr{\left(\hat{S}_{IL}^z +  \hat{S}_{IR}^z\right)^2~\hat{\varrho}_{\rm probe}}$ which is related to the probe magnetic susceptibility.

Since the probe reduced state eigenvalues $\varrho_k\equiv \varrho_k(\mathcal{C},\mathcal{M},\mathcal{\chi})$ are fully determined by these observables, the QFIM is clearly also a function of these quantities. Importantly, the factorized form of Eq.~\eqref{eq:qfim_factor_gen} no longer applies, and the QFIM singularity is removed due to the known control field. We find that,
\begin{equation}
\begin{split}
	&\varrho_{S} ={} \frac{1}{4} - \mathcal{C}\;,~~~~~~~~~~~~~~~
	\varrho_{T;0} ={} \frac{3}{4} + \mathcal{C} - \mathcal{\chi}\;,\\ 
	&\varrho_{T;-1} ={} \frac{1}{2}\left[\mathcal{\chi}- \mathcal{M} \right]\;,~~~
	\varrho_{T;+1} ={} \frac{1}{2}\left[\mathcal{\chi} + \mathcal{M} \right]\;. \label{eq:rhos}
\end{split}
\end{equation}
 We may now use Eq.~\eqref{eq:QFI_MP} to obtain the QFIM in terms of this set of observables,
\begin{equation}
    \label{Eq:QFIM_3_Obs}
	\begin{split}
\boldsymbol{\mathcal{H}}_{\lambda,\lambda'} =& 
\frac{\partial_{\lambda}(\mathcal{\chi}+\mathcal{M})\partial_{\lambda'}(\mathcal{\chi}+\mathcal{M})}{2(\mathcal{\chi}+\mathcal{M})}
+\frac{\partial_{\lambda}(\mathcal{\chi}-\mathcal{M})\partial_{\lambda'}(\mathcal{\chi}-\mathcal{M})}{2(\mathcal{\chi}-\mathcal{M})}\\ 
&+\frac{\partial_{\lambda}(\mathcal{C}-\mathcal{\chi})\partial_{\lambda'}(\mathcal{C}-\mathcal{\chi})}{3/4+\mathcal{C}-\mathcal{\chi}} +\frac{\partial_{\lambda}\mathcal{C} \partial_{\lambda'}\mathcal{C}}{1/4 - \mathcal{C}} 
	\end{split}
\end{equation}
In general, a finite magnetization $\mathcal{M}\ne 0$ is enough to remove the QFIM singularity. However, we note in passing that if the field $B$ is also unknown, then the $3\times 3$ QFIM for $T,K,B$ can once again become singular. The added information about the known field $B$ is crucial to this multiparameter estimation protocol. 

In Appendix~\ref{Appendix:Ansatz} we provide an approximate ansatz for the probe reduced state populations, which holds exactly in the large-$K$ and small $B$ limits, and is generally found to be rather accurate throughout the phase diagram. The simplification allows us to express the QFIM in terms of only $\mathcal{C}$ and $\mathcal{M}$. This might be advantageous in an experimental setting since simultaneous measurements of different observables may be challenging in practice. In principle, since the SLD operators for $T$ and $K$ commute, there exists a single optimal measurement basis (effectively the Bell basis on the probe states). However, this is not a physically meaningful or feasibly measurable observable.


\subsection{Multiparameter estimation in the large-$K$ limit}
\label{MP_Est}
As a simple but nontrivial demonstration of multiparameter estimation, we consider now the large-$K$ limit of the 2IK model with an applied control field $B$, focusing our discussion on the multiparameter QSNR $\mathcal{Q}_{MP}\left(\lambda,\lambda\right)$ for $\lambda = T$ or $K$. 

\begin{figure}[t]
    \centering
    \includegraphics[width=1\linewidth]{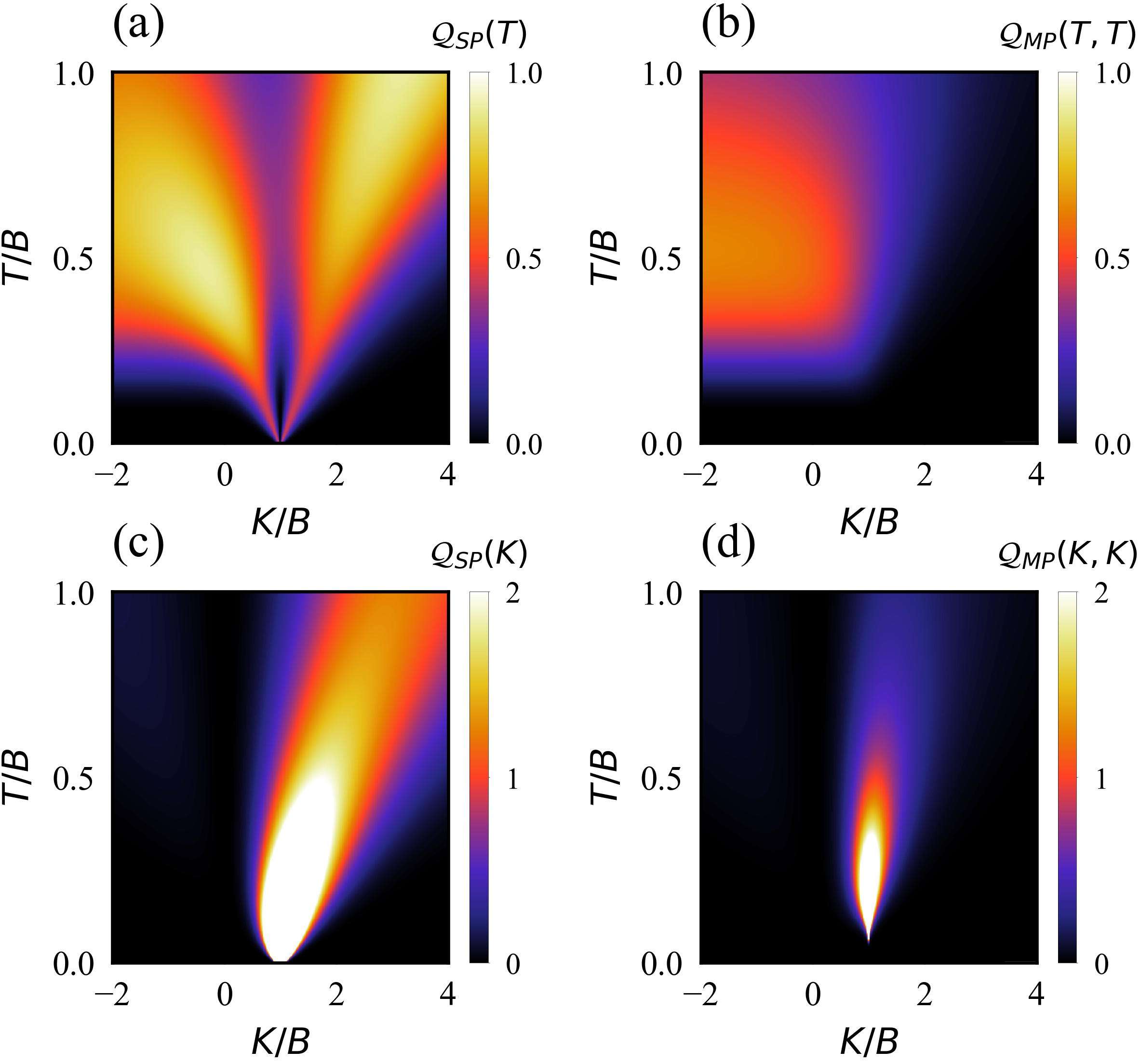}
    \caption{\textbf{Multiparameter estimation in the large-$K$ limit of the 2IK model.} (a) Single-parameter and (b) multiparameter QSNRs for thermometery. (c) Single-parameter and (d) multiparameter QSNRs for estimation of the coupling $K$. The multiparameter QSNR $\mathcal{Q}_{MP}(\lambda,\lambda)$ is a universal scaling function of $T/B$ and $K/B$, such that the regime of good sensitivity shrinks as the field strength is reduced, and vanishes in the limit $B\to 0$.}
    \label{Fig:KLIM_B_TK}
\end{figure}

By way of comparison, we consider first the analogous single parameter QSNRs $\mathcal{Q}_{SP}\left(\lambda\right)$, shown in Fig.~\ref{Fig:KLIM_B_TK}(a,c). See Fig.~\ref{Fig:KLIM_NoB_Thermo}(c,d) for the corresponding zero field case discussed already. The impact of the field itself on single-parameter estimation is discussed in depth in Appendix~\ref{sec:append_singlePcontrol} and \ref{sec:subopt}.

From Eq.~\eqref{Eq:MP_Thermo} we see that the QSNR for either $T$ or $K$, in the case where neither $T$ or $K$ are known with certainty, involves information from the full QFIM and not just its diagonal components. The behavior of the QFIM elements themselves are discussed in Appendix~\ref{Appendix:QFIM_Elements}. 
We note that $\mathcal{Q}_{MP}(\lambda,\lambda)$ and the quantity  $B^2\boldsymbol{\mathcal{H}}_{\lambda,\lambda}$ are universal scaling functions of the rescaled parameters $t=T/B$ and $k=K/B$ in the large-$K$ limit of the 2IK model. For example,
\begin{equation}
    \mathcal{Q}_{MP}\left(T,T\right) = \frac{2 e^{2/t}\left(2 + \cosh{1/t}\right)}{\left(1 + e^{1/t} + e^{2/t} \right)\left(1 + e^{1/t}\left(1 + e^{1/t} + e^{k/t}\right) \right)t^2}
    \end{equation}
and similarly for $\mathcal{Q}_{MP}\left(K,K\right)$. We present these universal results as a function of $t$ and $k$ in Fig.~\ref{Fig:KLIM_B_TK}. Interestingly,  $\mathcal{Q}_{MP}(\lambda,\lambda)$ and  $B^2\boldsymbol{\mathcal{H}}_{\lambda,\lambda}$ are also  universal functions of $t$ and $k$ in the full 2IK model provided all model parameters (in this case $j=J/B$ and $d=D/B$) are similarly scaled, although of course many-body effects change the functional form.

Turning now to our numerical results, we see a stark difference between the multiparametric QSNR for thermometry
$\mathcal{Q}_{MP}\left(T,T\right)$ shown in Fig.~\ref{Fig:KLIM_B_TK}(b) compared with its single-parameter estimation counterpart $\mathcal{Q}_{SP}\left(T\right)$ shown in 
Fig.~\ref{Fig:KLIM_B_TK}(a). Uncertainty in $K$ dramatically affects our ability to estimate $T$. In particular, $\mathcal{Q}_{MP}\left(T,T\right)$ is small for all $K>B$ and rapidly attenuates to zero at large $K$. Appreciable sensitivity to $T$ is instead afforded for $K<B$ because here the triplet state $|T;-1\rangle$ is the ground state, and excited states $|T;0\rangle$ and $|T;+1\rangle$ are respectively $B$ and $2B$ higher in energy. Since the control field strength $B$ is known, thermometric sensitivity is boosted at temperatures on the order of these excitation energies, since then the rate of population transfer is maximal. There is effectively no sensitivity at temperatures below the minimum gap $T\ll B$. The singlet state $|S\rangle$ is of order $|B-K|$ higher in energy, but $K$ is unknown. Therefore it is the triplet excitations which dominate thermometry in this system. This behavior is not evident from simply looking at the QFI for $T$, i.e. the element $\boldsymbol{\mathcal{H}}_{T,T}$ in the QFIM, as discussed in Appendix~\ref{Appendix:QFIM_Elements}.

We show the complementary behavior when estimating the coupling $K$ in the presence of uncertainty in the temperature $T$ in Fig.~\ref{Fig:KLIM_B_TK}(d) where we see the QSNR $\mathcal{Q}_{MP}\left(K,K\right)$ flares sharply for $K\sim B$ at the level crossing point between ground states, thus showing a more consistent behavior with the single-parameter estimation shown in panel (c). The lack of low-temperature thermometric sensitivity at $K=B$ means that estimation of $K$ near $K\sim B$ at low temperatures is also poor.


\subsection{Multiparameter estimation near criticality}
Richer behavior is naturally expected in the full 2IK model near the quantum critical point. The magnetic field $B$ is a relevant perturbation that also destabilizes the critical point. For small $B$ in the close vicinity of the critical point, the physics is again controlled by a single emergent scale $T^*$, to which all relevant perturbations contribute additively \cite{sela2011exact,andrew_entropy_exact}, $T^* \sim c_K \delta K^2 + c_B B^2$ with constants $c_K$ and $c_B$ of order $T_K^{-1}$. This expression replaces Eq.~\eqref{Eq:T_star_Def}, but Eqs.~\eqref{eq:entropy1}, \eqref{eq:entropy2} still hold. Therefore similar methods leading to Eqs.~\eqref{eq:Exact_Thermo_KC}, \eqref{eq:Exact_Kermo_KC} may be applied to obtain exact results for the multiparameter QSNRs. The generalization requires expressions for $\mathcal{M}$ and $\mathcal{\chi}$, but these can again be obtained from the entropy via the appropriate Maxwell relations by noting that $\mathcal{M}=\partial_B \mathcal{F}$ and $\mathcal{\chi}=\partial_B^2 \mathcal{F}$. We leave detailed calculations for future work.

In the full 2IK model our basic conclusion remains unchanged: multiparameter estimation is much more complex and subtle than its single-parameter counterpart. In particular, it is important to understand when a system might have a singular QFIM and how to remove that singularity. It is also crucial in assessing metrological performance to examine the QSNR rather than simply elements of the QFIM, which can paint a misleading picture in practice.


\section{Conclusions and Discussions}\label{Sec:Conc}

Quantum criticality is a well-known resource for quantum metrology at zero temperature. In practice, however, thermal fluctuations are inevitable and even the temperature might itself be unknown. In this work, we considered the 2IK model as a rich playground to explore finite-temperature multiparameter estimation in an  interacting quantum many-body system that supports a nontrivial second-order quantum phase transition. We regard the two spin-$\tfrac{1}{2}$ impurities as \textit{in-situ} probes for the continuum electronic environment in which they are embedded. Quantum sensing is performed by measuring physically-motivated observables on the accessible impurity qubit probes. The 2IK model and its variants can be realized in nanoelectronics quantum dot devices. Experimental signatures of finite-temperature quantum criticality in related Kondo systems have been recently reported \cite{potok2007observation,iftikhar2018tunable,pouse2023quantum}, but their metrological capacity was not  addressed.

Through the NRG method, we have computed the finite-temperature reduced density matrix of the two impurities as our probe state, and extracted from it elements of the QFIM for estimation of the system temperature $T$ and the probe coupling strength $K$. Interestingly, information on $T$ and $K$ appear only in the eigenvalues of the density matrix, which makes the optimal measurement basis independent of these parameters. As a novel figure of merit for quantifying the performance of the probe, we introduced the multiparameter QSNR. 

Remarkably, we were able to find exact analytic results for the QSNRs near the 2IK critical point at finite temperatures by relating them to physical observables on the probe, which were in turn obtained via Maxwell relations from an exact solution for the probe entropy.

In the context of single parameter estimation, our analysis shows that one can get strongly enhanced sensitivity (i.e. a large QSNR which diverges as $T \to 0$) for the coupling $K$ around the critical point at $K_c$ and up to the Kondo temperature $T_K$. High quality sensitivity to $K$ is also achieved outside the critical regime for $T \gg T_K$ -- albeit comparatively diminished with respect to achievable sensitivity near the critical point. This shows that criticality-based quantum sensors can indeed be beneficial even at finite temperatures. We also demonstrated that the probe can act as an effective thermometer, however only when the temperature $T$ is larger than the Kondo temperature $T_K$.

In the context of two-parameter estimation, where both the impurity coupling and the temperature are unknown, we find that the QFI matrix becomes singular, thus preventing estimation of either parameter. This is a counterintuitive result since QFI matrix singularities are typically associated with the interdependence of parameters to be estimated. Here by contrast, the parameters $T$ and $K$ are strictly independent, yet the QFI matrix is still singular. The reason for this is the partial accessibility to the system, which restricts the probe state to only the reduced density matrix of the two impurities. In order to restore metrological sensitivity, we show that simply applying a known control field is sufficient to remove the QFI matrix singularity.
Thanks to this simple recipe, the QFI matrix becomes invertible and simultaneous two parameter estimation becomes possible. Our analysis shows that the precision of estimating any parameter is enhanced if the other parameters are known. If this is not possible, the precision can still be strongly enhanced even if another parameter is unknown, but its QFI is large. Since uncertainties are unavoidable in any practical situation, this provides a conceptual route to the design of optimized and robust quantum sensors: one should seek to maximize the QFI of \emph{all} model parameters -- not just the one directly of interest for metrology.  

Our results provide important insights for probe-based parameter estimation of complex quantum systems, in particular highlighting the importance of the QSNR as a key figure of merit to accurately assess a system's metrological utility. The 2IK model provides a versatile setting to demonstrate our results; however we expect that similar effects will be relevant to other candidate systems for critical quantum metrology.


\begin{figure*}
	\begin{center}
		\includegraphics[width=1.\linewidth]{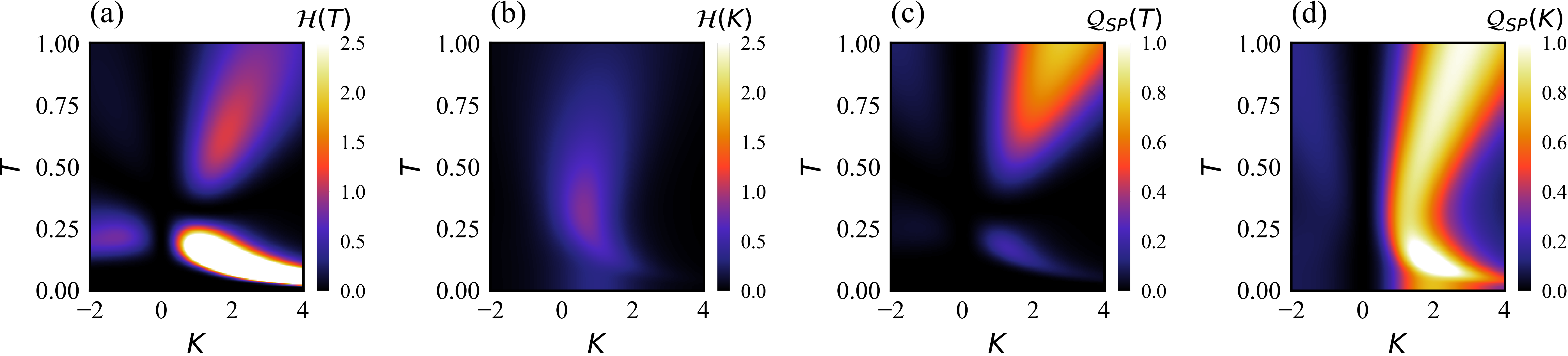}
	\end{center}
	\caption{\textbf{Single-parameter estimation of $T$ and $K$ in the narrow band limit of the 2IK model.}  \emph{Panels a-b:} QFI for thermometry $\mathcal{H}\left(T\right)$ and for the estimation of probe coupling strength $\mathcal{H}\left(K\right)$ as a function of $T$ and $K$. \emph{Panels c-d:} Corresponding single parameter QSNR responses $\mathcal{Q}_{SP}\left(T\right)$ and $\mathcal{Q}_{SP}\left(K\right)$ as a function of $T$ and $K$. Results obtained by exact diagonalization.}
	\label{Fig:NBL_Results_Appendix}
\end{figure*}

\acknowledgements
We acknowledge fruitful exchanges with Mark Mitchison and Gabriel Landi. GM acknowledges support from Equal1 Laboratories Ireland Limited. SC is supported by the Science Foundation Ireland Starting Investigator Research Grant “SpeedDemon” No. 18/SIRG/5508, the John Templeton Foundation Grant ID 62422, and the Alexander von Humboldt Foundation. AKM acknowledges funding from the Irish Research Council through grant EPSPG/2022/59 and from Science Foundation Ireland through grant 21/RP-2TF/10019. AB acknowledges support from the National Natural Science Foundation of China (Grants Nos. 12050410253, 92065115, and 12274059) and the Ministry of Science and Technology of China (Grant No. QNJ2021167001L).

\appendix

\section{Uncertainties in multiparameter estimation}\label{Appendix:Uncertainties_multiparameter}

One can multiply both sides of the multiparameter Cram\'er-Rao inequality inequality, given in Eq.~(\ref{eq:MPQCRB}), by a positive weight matrix $W\ge 0$ and take the trace of both sides to get an inequality for scalar quantities, namely
\begin{equation}
\label{eq:MPQCRB-weighted}
\textrm{Tr} \left[ W \textbf{Cov}\left[\vec{\lambda}\right] \right] \geq \frac{1}{N}\textrm{Tr} \left[ W\boldsymbol{\mathcal{H}}^{-1} \right].
\end{equation}
If one is only interested in estimating one of the parameters, i.e. $\lambda_i$, then we can choose $W=\textrm{diag}(0,\cdots,0,1,0,\cdots,0)$, where only $W_{ii}=1$. In this case the Eq.~(\ref{eq:MPQCRB-weighted}) becomes
$\textrm{Var}(\lambda_i) \geq \frac{1}{N} (\boldsymbol{\mathcal{H}}^{-1})_{ii}$.
Note that here the other parameters are assumed to be unknown. In the case that all other parameters are known, the problem reduces to a single parameter estimation in which $\textrm{Var}(\lambda_i) \geq \frac{1}{N} (\boldsymbol{\mathcal{H}}_{ii})^{-1}$. The QFI matrix is positive semi-definite, i.e. $\boldsymbol{\mathcal{H}}\geq 0$, which implies that $(\boldsymbol{\mathcal{H}}^{-1})_{ii} \geq (\boldsymbol{\mathcal{H}}_{ii})^{-1}$. This clearly shows that the uncertainty in estimating $\lambda_i$ reduces if one knows other parameters of the system. 


\begin{figure*}[t]
	\begin{center}
		\includegraphics[width=1.\linewidth]{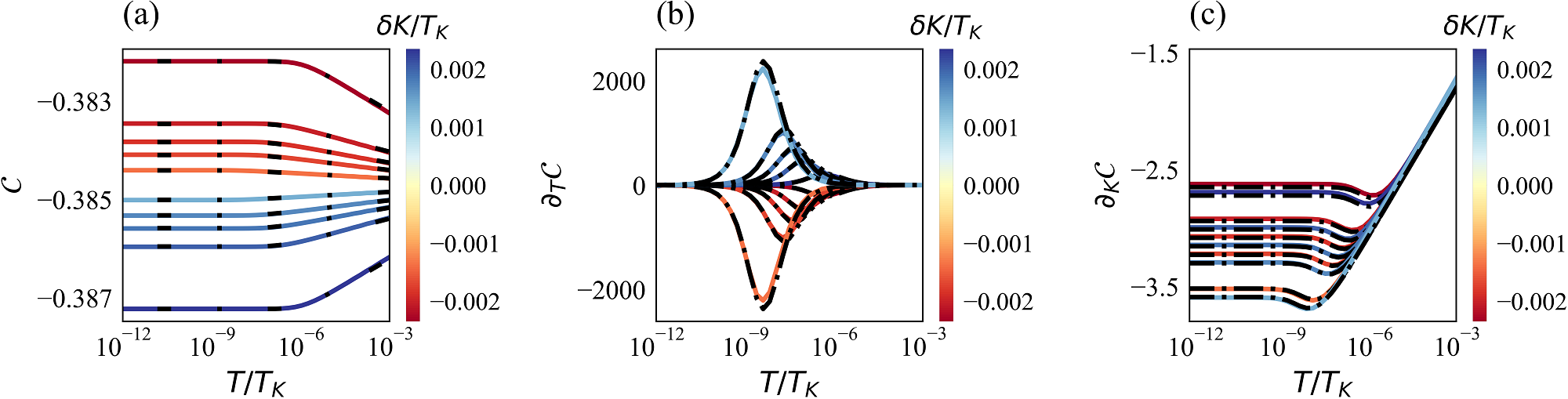}
	\end{center}
	\caption{Exact analytic results for the spin-spin correlation function $\mathcal{C}$ and its derivatives $\partial_T \mathcal{C}$ and $\partial_K \mathcal{C}$ (colored lines), compared with numerically-exact NRG results (black dashed lines) in the universal quantum critical regime of the full 2IK model. Plotted for $J=1$.}
	\label{Fig:Exact_Results_Validation_Appendix}
\end{figure*}

\section{Narrow band limit}\label{Appendix:Narrow_Band_Limit}
In Sec.~\ref{Sec:large_K} of the main text, we consider an analytically tractable limit of the full 2IK system where the fermionic leads play no role in determining the reduced state of the probe. Another important limit is that arising when $J/D\gg 1$ (with $D$ the conduction electron bandwidth), wherein the continuum of electronic states in the leads can be well-approximated by just the single orbital in real-space to which each impurity couples (previously denoted $\hat{c}_{\alpha 0 \nu}$). This is known as the `narrow band limit' (NBL), and is described by the Hamiltonian,
\begin{equation}
	\label{Eq:NBL_Ham}
	\hat{H}_{NBL} = K~\hat{\vec{\mathbf{S}}}_{IL}\cdot\hat{\vec{\mathbf{S}}}_{IR} + J~\left ( \hat{\vec{\mathbf{S}}}_{IL}\cdot\hat{\vec{\mathbf{s}}}_{EL} + \hat{\vec{\mathbf{S}}}_{IR}\cdot\hat{\vec{\mathbf{s}}}_{ER} \right )
\end{equation}
The competition between $J$ and $K$ scales is captured in the NBL, although many-body effects (such as Kondo and the quantum phase transition) are missing. The two-impurity probe RDM is still a function of the single observable $\mathcal{C}$, but it is no longer just a function of the single reduced parameter $K/T$ since we also now have a nontrivial dependence on $J/T$. These properties have implications for metrology, as shown below.

The NBL, which features just one electronic mode in each bath as per Eq.~\eqref{Eq:NBL_Ham}, is nonetheless a valid description of the system when $J\gg D$. This limit has the advantage that it is simple enough to be analytically tractable, like the large-$K$ limit (although in practice we do not present closed-form formulae here since they are extremely cumbersome). For concreteness in this section we set $J=1$ and explore the $(T,K)$ parameter space as before.

We stress that, although the physical setup includes the explicit addition of environment degrees of freedom, the two spin-$\tfrac{1}{2}$ impurity composite probe is still fully characterised by the spin-spin correlation function $\mathcal{C}$, and metrological properties can be extracted from Eq.~\eqref{eq:QFI-SP-C}. We note that explicit inclusion of the impurity-environment coupling can significantly alter the physics of the system, as manifest in the behavior of $\mathcal{C}$. This is because the impurity probes can now become entangled with the environment degrees of freedom. 
Unlike the large-$K$ limit, in the NBL the overall ground state of the system is always a net spin-singlet state and there is no ground state transition as a function of intra-probe coupling strength $K$. However, the character of the reduced state on the two-impurity probe changes continuously when $J$ is finite from being predominantly singlet type on the impurity probes for $K>0$ to predominantly triplet type for $K<0$. This is clearly demonstrated through the smooth evolution of the impurity probe spin-spin correlator $\mathcal{C}$ with coupling strength $K$ (not shown), which in turn controls metrological performance in this system. Therefore we do again expect qualitative differences between positive and negative $K$ regimes, but a smoother crossover in behavior around $K=0$. We also note however that the NBL does support transitions (level crossings) in the excited states, which play a role in finite-temperature metrology. We expect the finite $J$ scale to shift the maximum measurement precision point to finite $K$ and $T$ (by contrast, in the large-$K$ limit the maximum QFI was somewhat artificially pinned to $K=T=0$).

We now turn to our numerical results for single-parameter estimation in the NBL of the 2IK model in panels (a-d) of Fig.~\ref{Fig:NBL_Results_Appendix}. A first glance at the QFI plots shows a richer structure compared with the large-$K$ limit shown in the main text in panels (a-d) of Fig.~\ref{Fig:KLIM_NoB_Thermo}. We see that the maximum in both $\mathcal{H}(T)$ and $\mathcal{H}(K)$ occurs around $K\sim J$ at finite temperatures. More instructive are the QSNR responses in Figs.~\ref{Fig:NBL_Results_Appendix}(c,d). First, we observe that the QSNR for estimation of $T$ and $K$ are now different from each other. We note that for large $K\gg J$ we start to recover the large-$K$ limit behavior in panels (c,d) of Fig.~\ref{Fig:KLIM_NoB_Thermo}, especially at higher temperatures. However, the low-$T$ behavior around $K\sim J$ is dramatically altered, with a fragmentation of the thermometric response $\mathcal{Q}_{SP}(T)$ to a double-peak structure, and a pronounced knee in $\mathcal{Q}_{SP}(K)$. These features anticipate what will become an actual quantum phase transition in the full many-body model. The results seem to show that with coupling to the leads, good thermometric sensitivity is not possible at low temperatures $T\ll J$. By contrast, the peak sensitivity in estimating $K$ is at around $K\sim J$. For thermometry we explicitly note that only at larger $T$ and $K$ do we see good temperature estimation capability due to finite $J$ scale - this is rather similar to the full 2IK model shown in panel (g) of Fig.~\ref{Fig:KLIM_NoB_Thermo}. Similarly, the single-parameter estimation of the coupling constant $K$ becomes relatively enhanced when $K$ is of order $J$ which anticipates the onset of criticality.


\section{Validation of exact result near criticality}\label{Appendix:Validation_Exact_Result}

In Sec.~\ref{sec:exact_results} of main text we present exact analytical results for the spin-spin correlation function $\mathcal{C}\left(T,K\right)$ in Eq.~\eqref{eq:corr}, as well as its derivatives $\partial_T\mathcal{C}\left(T,K\right)$ in Eq.~\eqref{eq:Exact_C} and $\partial_K\mathcal{C}\left(T,K\right)$ in Eq.~\eqref{eq:exact_c_k}, for the full 2IK model in the close vicinity of the critical point. In turn, this allows us to find closed form solutions for both the thermometric precision $\mathcal{Q}_{SP}^2\left(T\right)$ in Eq.~\eqref{eq:Exact_Thermo_KC} and sensitivity to coupling strength $\mathcal{Q}_{SP}\left(K\right)$ in Eq.~\eqref{eq:Exact_Kermo_KC}. In this section we validate these analytic expressions by comparison with full NRG calculations. For concreteness we carry out our NRG calculations with a coupling $J = 1$, for which we find that $K_C \simeq 0.618$, $T_K = 0.362$, $\mathcal{C}^* \simeq - 0.385$ and we extract the constant $c \simeq 0.035$ from NRG thermodynamics. 

In panel (a) of Fig.~\ref{Fig:Exact_Results_Validation_Appendix} the NRG computed spin-spin correlation function is plotted for different values of impurity detuning, $\delta K = K - K_C$, where blue and red lines correspond to positive and negative detunings $\delta K$ respectively. The corresponding exact results of Eq.~\eqref{eq:corr} are denoted by the black dashed line. We see excellent agreement between the NRG and exact solutions, and note that the spin-spin correlation function remains largely unchanged as a function of $T$ at low temperatures, but changes sharply as a function of the impurity detuning $\delta K$. Similarly, in panels (b) and (c) of Fig.~\ref{Fig:Exact_Results_Validation_Appendix} we show the spin-spin correlation function derivatives, $\partial_T\mathcal{C}\left(T,K\right)$ and $\partial_K\mathcal{C}\left(T,K\right)$ respectively in the colored red and blue lines for the NRG solution, and the black dashed lines shows the derived exact results Eq.~\eqref{eq:Exact_C} and Eq.~\eqref{eq:exact_c_k}. We again note the excellent quantitative agreement. 


\begin{figure*}[t]
	\begin{center}
		\includegraphics[width=1.\linewidth]{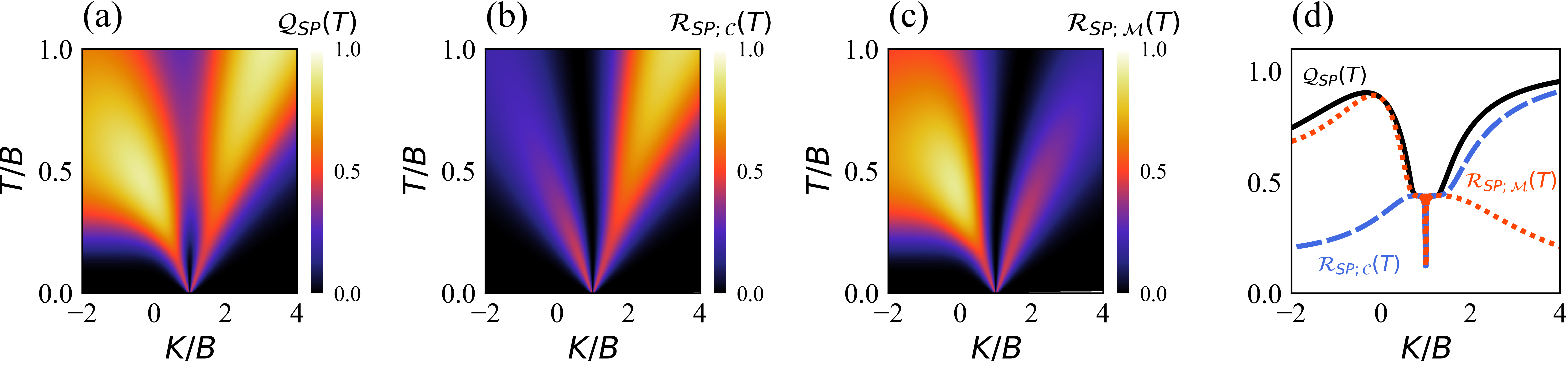}
	\end{center}
	\caption{\textbf{Sub-optimal thermometry in the large-$K$ limit of the 2IK model with an applied field $B$.}  (a) Single-parameter quantum signal to noise ratio, $\mathcal{Q}_{SP}\left(T\right)$ as a function of rescaled temperature $T/B$ and coupling strength $K/B$; compared with (b) the (sub-optimal) single-parameter signal to noise ratio $\mathcal{R}_{SP;\mathcal{C}}\left(T\right)$ using the Fisher information obtained from measurement of only the spin-spin correlation function $\mathcal{C}$; and (c) the (sub-optimal) signal to noise ratio $\mathcal{R}_{SP;\mathcal{M}}\left(T\right)$ corresponding to a magnetization measurement $\mathcal{M}$. The maximum QSNR and SNRs are shown in (d) for a given $K/B$. Panel (a) for the QSNR with finite field should be compared with the $B=0$ result shown in Fig.~\ref{Fig:KLIM_NoB_Thermo}(c).}
	\label{Fig:Suboptimal_Appendix}
\end{figure*}

\section{Approximate ansatz for the probe reduced\\state in a $B$ field}\label{Appendix:Ansatz}
The situation described in Sec.~\ref{Control_Field}, which is formally exact, can be simplified somewhat by making an approximate ansatz for the probe reduced state populations. We motivate this by looking at the large-$K$ limit of the 2IK model, where we take the probe reduced density matrix to be that of the isolated probe, thermalized to the same temperature as the environment. In that case, we may write $\varrho_{T;0}\equiv \varrho_T$ and $\varrho_{T;+1}=\varrho_T\times e^{-B/T}$, $\varrho_{T;-1}=\varrho_T\times e^{+B/T}$ due to the energy splitting of the spin-triplet states in a field. With explicit coupling to the leads (finite $J$), this is no longer the case because the probe and leads become entangled (the full states are not probe-lead separable). However, one might expect that the $S^z=\pm 1$ reduced states of the probe are still split approximately symmetrically around the $S^z=0$ triplet state, with some renormalized effective field. Taking the ansatz $\varrho_{T;+1}/\varrho_{T;0}=\varrho_{T;0}/\varrho_{T;-1}$, which holds exactly in the large-$K$ limit and also in the small-$B$ limit, we can derive an expression for  $\mathcal{\chi}$ in terms of $\mathcal{C}$ and $\mathcal{M}$,
\begin{equation}\label{eq:chi}
    \mathcal{\chi}\simeq  1+\tfrac{4}{3}\mathcal{C}-\sqrt{(\tfrac{1}{2}+\tfrac{2}{3}\mathcal{C})^2-\tfrac{1}{3}\mathcal{M}^2}
\end{equation}
This allows us to eliminate the dependence of $\mathcal{\chi}$ and reduce the parametric complexity of the problem to just two observables -- the probe spin-spin correlator $\mathcal{C}$ and the magnetization $\mathcal{M}$. Even in the full 2IK model with finite $K,J,B,D$ we find from our NRG calculations that Eq.~\eqref{eq:chi} holds rather accurately across the full parameter space investigated. 


\section{Single-parameter thermometry with a control field}\label{sec:append_singlePcontrol}
In Sec.~\ref{Sec:MultiParam_Background} we apply a known control field $B$ to our system which removes the singularity of the QFIM. Here we briefly discuss how a finite field $B$ alters the physics of the system by summarizing results for the single-parameter estimation case, focusing on the QSNR for thermometry in the large-$K$ limit of the 2IK model. The situation here is somewhat different from that of  Fig.~\ref{Fig:KLIM_NoB_Thermo}(c) due to the inclusion of the field. In Fig.~\ref{Fig:KLIM_B_TK}(a) of the main text we plot $\mathcal{Q}_{SP}(T)$ as a function of the rescaled parameters $t=T/B$ and $k=K/B$. The single-parameter QSNR is a universal function of $t$ and $k$. This holds exactly in the large-$K$ limit, but also holds generally in the full 2IK model provided all parameters are similarly scaled. This scaling implies that we get the same behavior as a function of $t$ and $k$ independent of the field strength $B$ (but over an ever-narrowing window of the bare parameters $T$ and $K$ as $B\to 0$). 

The single-parameter QSNR in Fig.~\ref{Fig:KLIM_B_TK}(a) of the main text shows level-crossing transition behavior at $K\!=\!B$ corresponding to a change in ground state from the collective probe spin-singlet $|S\rangle$ for $K>B$ to a component of the spin-triplet $|T;-1\rangle$ for $K<B$. The transition between singlet and triplet ground states occurs at $K=0$ when $B=0$, as shown in Fig.~\ref{Fig:KLIM_NoB_Thermo}(c). However, at $B=0$ we saw that the transition is also associated with a change in the degeneracy of the ground and excited states, and this produced a marked difference in the maximum attainable QSNR for $K>0$ vs $K<0$. At finite $B$ and modest $K$ we see no such difference between the phases $K>B$ and $K<B$ because all states are unique (non-degenerate) due to splittings in the field. At very large $|K/B|$ and $K>0$ we expect to recover the QSNR profile obtained at $B=0$, with boosted sensitivity around $T\sim K$ as the excited triplet states become quasi-degenerate. For large $|K/B|$ but $K<0$, we have relatively poor sensitivity at $T\sim K$ because the excited singlet state is non-degenerate, but transitions between the triplet states now provide additional sensitivity for $T\sim B$.


\begin{figure*}[t]
	\begin{center}
		\includegraphics[width=.9\linewidth]{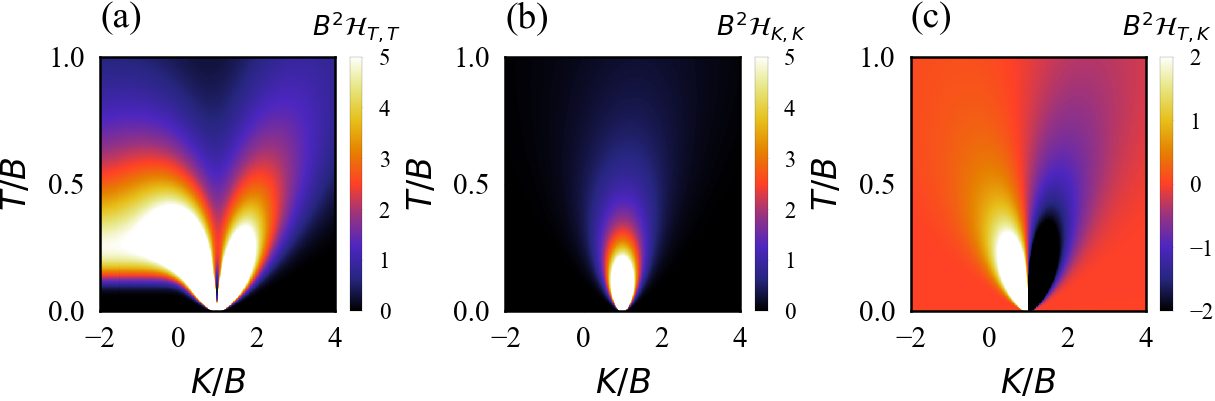}
	\end{center}
	\caption{\textbf{Elements of the multiparameter QFI matrix in the large-$K$ limit of the 2IK model.} We plot $B^2 \boldsymbol{\mathcal{H}}_{T,T}$, $B^2 \boldsymbol{\mathcal{H}}_{K,K}$, and $B^2 \boldsymbol{\mathcal{H}}_{T,K}$ ($\equiv B^2 \boldsymbol{\mathcal{H}}_{K,T}$) in panels (a), (b), and (c) respectively.  The quantities $B^2 \boldsymbol{\mathcal{H}}_{\lambda,\lambda'}$ are universal scaling functions of $T/B$ and $K/B$.}
	\label{Fig:Multi_Param_QFI_Appendix}
\end{figure*}

\section{Sub-optimal measurements}\label{sec:subopt}
The QSNR plotted in Fig.~\ref{Fig:KLIM_B_TK}(a) of the main text and in Fig.~\ref{Fig:Suboptimal_Appendix}(a) is the `best case scenario' obtained from the (single parameter) thermometric QFI -- it corresponds to the optimal measurement on the probe. This optimal measurement (the thermometric SLD) always exists and is an observable in principle. For the 2IK model, the optimal measurement basis is effectively the Bell basis, but such measurements are not practical in an experimental setting. Previously, we have assumed that we have simultaneous access to the set of physical observables $\mathcal{C}$, $\mathcal{M}$ and $\mathcal{\chi}$, in terms of which the SLD can be decomposed. Here we ask about the thermometric sensitivity obtainable for a sub-optimal measurement -- for example just the spin-spin correlator $\mathcal{C}$ or the magnetization $\mathcal{M}$. With only partial access to the probe state, one naturally expects lower parameter estimation precision.

This can be understood quantitatively from the Fisher information $F_{\Omega}(T)$ for a given measurement $\Omega$. The corresponding (sub-optimal, single-parameter) signal-to-noise ratio (SNR) is defined analogously to the QSNR,
\begin{equation}
	\mathcal{R}_{SP:\Omega}\left(T\right) = T^2~F_{\Omega}\left(T\right) \;.
\end{equation}
with $F_{\Omega}\left(\lambda\right) = \abs{\partial_{\lambda}\expval{\Omega}}^2/\text{Var}\left[\Omega \right]$ and $\lambda = T$ for thermometry, defined from the error propagation formula \cite{toth2014quantum}. 
In Fig.~\ref{Fig:Suboptimal_Appendix}(b,c) we plot the thermometric SNR corresponding to measurement of just $\mathcal{C}$ or $\mathcal{M}$ on the probe (which are the dominant contributions). Fig.~\ref{Fig:Suboptimal_Appendix}(d) shows the maximum QSNR or SNR at a given $K/B$. We see immediately that the full QSNR can be decomposed into contributions from the different measurements (although not in a simple additive fashion, as Eq.~\eqref{Eq:QFIM_3_Obs} attests). In particular, note that the measurement of only $\mathcal{C}$ may yield better results than measuring just $\mathcal{M}$ in certain parameter regimes (in this case, for $K>B$) -- with the SNR $\mathcal{R}_{SP:\mathcal{C}}$ even approaching the QSNR $\mathcal{Q}_{SP}$ for certain parameters. However measurement of $\mathcal{M}$ might be preferable in other cases (for example $K<B$). The differences can be traced to the different types of transitions controlling the SNR in the two regimes: for $K>B$ the dominant contributions come from spin singlet-triplet transitions picked up by the spin-spin correlator $\mathcal{C}$, whereas for $K<B$ magnetic transitions between the triplet components dominate, and these sensitively affect the magnetization $\mathcal{M}$. Therefore intimate details about the spectrum can be inferred by examining the SNRs for different but complementary observables in this way. Broadly similar results are obtained for the single-parameter estimation of the coupling constant $K$ (not shown).\\
\vfill


\section{QFIM for multiparameter estimation}\label{Appendix:QFIM_Elements}
In the main text Sec.~\ref{MP_Est} we present the multiparameter QSNRs in the large $K$-limit in the presence of a magnetic field. In Fig.~\ref{Fig:Multi_Param_QFI_Appendix} we provide the elements of the QFIM for reference. In particular, we note that the off-diagonal element, $\boldsymbol{\mathcal{H}}_{T,K}$ presented in panel (c) can be negative if there is a negative correlation between the parameters; the variance or covariance does not correspond simply to any one of the QFIM elements, but requires information from all of them.

Formally, the diagonal terms of the QFIM are related to variances in the single-parameter estimation case, and so these quantities are strictly positive. On the other hand, the off-diagonal terms can be either positive or negative -- the sign telling us in what way the parameters are correlated.  

Our results illustrate the importance of using the QSNR rather than the QFIM itself to quantify metrological performance. For example, the QFIM element $\boldsymbol{\mathcal{H}}_{T,T}$ shown in Fig.~\ref{Fig:Multi_Param_QFI_Appendix}(a) only provides us with information about how temperature sensitivity is imprinted on our probe in the context of single parameter estimation. 
The true multiparameter QSNR $\mathcal{Q}_{MP}\left(T,T\right)$ shown in Fig.~\ref{Fig:KLIM_B_TK}(b) tells a very different story about the actual metrological utility of the probe.
\vfill

\bibliography{bibo}

\end{document}